\documentclass[a4paper,fleqn,usenatbib]{mnras}

\usepackage[T1]{fontenc}
\usepackage{ae,aecompl}

\usepackage{graphicx}	
\usepackage{amsmath}	
\usepackage{amssymb}	
\usepackage{hyperref}
\usepackage{pdflscape}	

\title[Colour jumps across the spiral arms of HUDF galaxies]
{Colour jumps across the spiral arms of Hubble Ultra Deep Field galaxies}

\author[Mart\'{\i}nez-Garc\'ia et al.]{
Eric E. Mart\'inez-Garc\'ia ,$^{1}$\thanks{E-mail: ericmartinez@inaoep.mx}
Rosa A. Gonz\'alez-L\'opezlira,$^{2}$,
and Iv\^anio Puerari.$^{3}$
\\
$^{1}$CONAHCYT Research Fellow~-~Instituto Nacional de Astrof\'isica, \'Optica y Electr\'onica, Luis E. Erro 1, Tonantzintla, \\ Puebla, C.P. 72840, M\'exico\\
$^{2}$Instituto de Radioastronom\'ia y Astrof\'isica, UNAM, Campus Morelia, Michoac\'an, M\'exico, C.P. 58089\\
$^{3}$Instituto Nacional de Astrof\'isica, \'Optica y Electr\'onica, Luis E. Erro 1, Tonantzintla, Puebla, C.P. 72840, M\'exico\\
}

\date{Accepted XXX. Received YYY; in original form ZZZ}

\pubyear{2022}

\begin{document}
\label{firstpage}
\pagerange{\pageref{firstpage}-~-\pageref{lastpage}}
\maketitle

\begin{abstract}

We have measured, at various wavelengths, the spiral arm pitch angles of 
a sample of distant spiral galaxies from the Hubble Space Telescope eXtreme Deep Field (XDF).
According to density wave theory, we should detect colour jumps from red-to-blue across the spiral arms.
Colour jumps are a consequence of large-scale shocks, which also generate the classic blue-to-red age/colour gradients,
and have only been detected until now in nearby spiral galaxies. Our results indicate that colour jumps
and gradients have been occurring in distant galaxies for at least the last 8 Gyr, in agreement with density wave theory.

\end{abstract}




\section{Introduction}

The origin of spiral structure in disc galaxies has been a matter
of debate over the last decades. The most recent reviews on the topic
include those of~\citet{dob14},~\citet{shu16}, and~\citet{sell20}, who arrive to different
conclusions.~\citet{dob14} argue that the long-lived waves envisioned by
density wave (DW) theory~\citep{lin64,ber89} are not produced in modern numerical simulations,
and hence that spiral structure may be only transient and recurrent, as shown by the simulations.
On the other hand,~\citet{shu16} argues in favour of long-lived spirals with a constant spiral pattern speed.
DW theory also foretells that large-scale shocks of gas induce star formation~\citep{rob69},
and consequently an age (or colour) gradient should be observed across the spiral arms,
downstream of the flow for regions inside the corotation radius ($R_{\mathrm CR}$),\footnote{
This is the radius where the angular velocity of the density wave, i.e., the pattern speed $\Omega_{\rm p}$,
is equal to the angular velocity of material in the disc, $\Omega(R)$.}
and upstream of the flow for regions outside $R_{\mathrm CR}$. There is plenty of observational
evidence supporting this scenario~\citep[e.g.,][]{gon96,pue97,egu04,tam08,egu09,gros09,mart09a,mart09b,mart11,san11,mart13,ced13,mart14,yu18,pet19,vall21a,vall21b,kar22}.
Spiral arms also seem to be amplitude-modulated, in accordance with DW theory~\citep{gros88,elm89,elm95,pue00}.
Furthermore,~\citet{sell20} discuss that spiral structure must be primarily a self-excited 
phenomenon in disc galaxies, and less frequently caused by tidal interactions or bars.
Externally excited spiral activity is not long-lived, as expected by DW theory,
but instead dissipates in about ten disc rotations~\citep[1-2 Gyr;][]{sell84,sell14}.
Spiral arm modes in disc galaxies may be maintained for several rotations if 
the disc is cooled by gas or other dynamical mechanism that prevents the Toomre parameter
from increasing its value, and consequently allows the non-asymmetric structure in
the disc to survive.~\citet{sell20}, however, acknowledge that there is no compelling evidence that
spiral arms in real galaxies are undoubtedly caused by the mechanisms suggested
by numerical simulations~\citep[with the exception perhaps of the Milky Way galaxy; see][]{sell19}.

In this paper we will focus on observations of spiral
galaxies beyond our nearby universe, in order to compare them with the predictions
of spiral structure formation theories and numerical simulations.

\subsection{Colour gradients across spiral arms}~\label{sect_colorgrads}

\begin{figure*} 
\centering
\includegraphics[angle=-90,width=0.8\hsize]{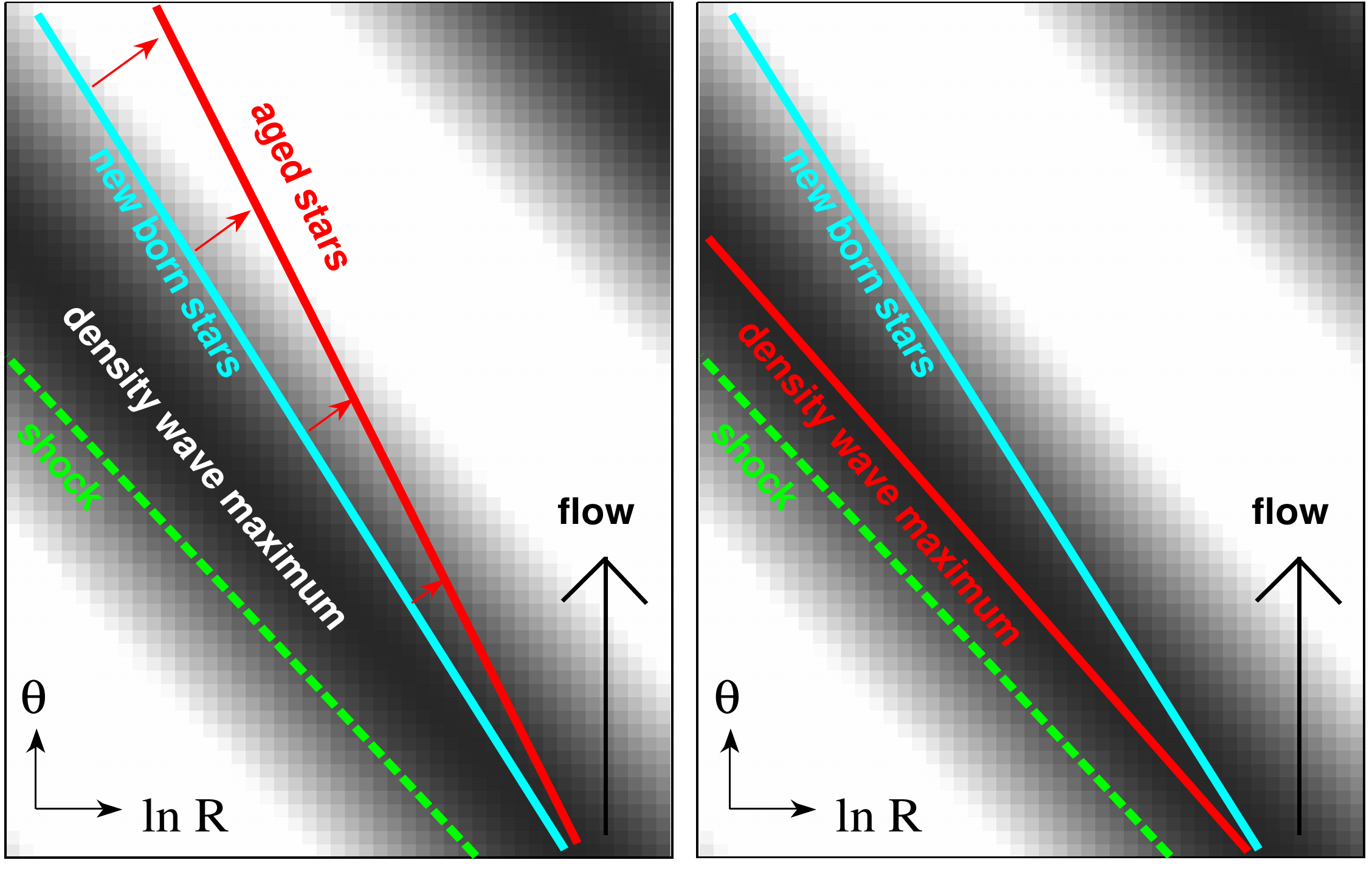}
\caption[fig1]{{\it Left panel}:
Classic age gradient across trailing spiral arms (blue-to-red),
in the $\ln R$ vs. $\theta$ plane, for radii $R < R_{\mathrm CR}$.
The spiral shock, density wave, newborn stars,
and aged stars are represented by the dashed green line, dark gray region, 
continuous blue line, and continuous red line, respectively.
{\it Right panel}:
Colour jump across trailing spiral arms (red-to-blue).
The spiral shock, density wave maximum, and new born stars,
are represented by the dashed green line, continuous red line, 
and continuous blue line, respectively.
The `colour jump' is located upstream of the classic age gradient (see left panel).
In this plot, a circle ($R$=constant) is represented by a vertical line and has a pitch angle of $P=0\degr$.
}~\label{fig1}
\end{figure*}

There are two opposite but coexisting colour gradients that can been observed across spiral arms~\citep[see, e.g.,][]{yu18}.
The first one actually envisioned by~\citet{rob69}, i.e., the
`classic' age (colour) gradient, is caused by the aging of newborn stars
as they drift away from the spiral arms. For trailing spirals,
the expected colour sequence is blue-to-red~(see Figure~\ref{fig1}, left panel).
This is the kind of gradient studied by~\citet{gon96} and~\citet{mart09a}, who
employ a reddening-free $Q$-index to trace aging newborn stars. Classic age gradients
are difficult to observe along the entire spiral arms due to masking by dust, radiation from HII regions,
and the substructure itself of the spirals arms~\citep[e.g.,][]{lav06,she06}.
Nevertheless, the statistics of `classic' age gradients agree with the theoretical expectations,
even when a circular model (i.e., one that neglects non-circular motions) is employed to determine $\Omega_{\rm p}$~\citep[see e.g.,][]{mart09b,mart15}.

The other `gradient' that can be detected across spiral arms
is a `colour jump' from red-to-blue~(see Figure~\ref{fig1}, right panel).
\citet{gitt04} predicted that the pitch angle\footnote{The angle between a tangent
to the spiral arm at a certain point and a circle, whose centre coincides
with the galaxy’s, that crosses the same point.} of the star-formation (SF-) arm, $P_{\rm SF}$ (traced by newly born stars),
should be smaller than the pitch angle of the potential minimum (or local maximum surface mass density, $P$-arm),
$P_{\rm pot}$ (traced by old stars, i.e., the density wave), such that $P_{\rm SF} < P_{\rm pot}$.
In this scenario, spiral arms at shorter wavelengths are expected to have a smaller pitch angle (more tightly wound arms).
The colour jump is due to the angular offset between the $P$-arm and the SF-arm, which is proportional
to the difference between the angular velocity of the stars and gas in the disc,
and the pattern speed, $\Omega - \Omega_{\rm p}$. Because this difference vanishes at
corotation, the angular offset between the two arms will also vanish, which requires $P_{\rm SF} < P_{\rm pot}$.
On the other hand, the shock (traced by the dust lanes and with the maximum gas density),
upstream of the $P$-arm, is expected to have a pitch angle, $P_{\rm shock}$,
equal to or smaller than $P_{\rm pot}$, but larger than $P_{\rm SF}$~\citep{gitt04}.
Colour jumps have been successfully detected by~\citet{gros98},
~\citet{mart13},~\citet{mart14b},~\citet{yu18}, and~\citet{line22}.

If density waves were already triggering star formation several Gyr ago,
we should be able to observe `colour jumps' across spiral
arms in galaxies with redshifts $z>0.1$. Other compelling theories,
such as self-propagating star formation~\citep{mue76,ger78},
or transient and recurrent arms~\citep{gra13,mich14}, where the spiral arm pitch
angle is mainly constrained by the shear rate of the galactic rotation curve,
do not predict any colour jumps or age (colour) gradients.

This paper is organized as follows.
In Section~\ref{data_sample} we present the galaxy sample,
in Section~\ref{analysis} we delineate the analysis,
in Section~\ref{results} we describe and discuss the results,
and finally in Section~\ref{conclu} we give our conclusions.


\section{Galaxy sample}~\label{data_sample}

The selection of our sample of galaxies was first made by inspecting the data products of the
{\it Hubble Space Telescope} ({\sl HST}) eXtreme Deep Field~\citep[XDF,][]{illi13}.
The XDF combines data sets from the original Hubble Ultra-Deep Field~\citep[HUDF,][]{bec06},
the HUDF09, the Cosmic Assembly Near-infrared Deep Extragalactic Legacy Survey~\citep[CANDELS,][]{gro11},
the HUDF12, and supernovae follow-ups. There are two data sets with a pixel scale of 30 milliarcseconds (mas), and 60 mas, respectively.
The 30 mas data set comprises five optical filters: F435W, F606W, F775W, F814W, and F850LP,
from the Advanced Camera for Surveys (ACS),
while the 60 mas data set includes the aforementioned optical ACS filters plus four Wide-Field Camera 3 (WFC3)
near-infrared (NIR) filters: F105W, F125W, F140W, and F160W.
The {\sl HST} point-spread function (PSF) full-width-at-half-maximum
(FWHM) is $\sim0\farcs1-0\farcs2$ (see Table~\ref{tbl-1}).


\begin{table}
	\centering
	\caption{PSF FWHM of the XDF data, obtained by measuring foreground stars in the 60 mas images.}
	\label{tbl-1}
	\begin{tabular}{cc} 
		\hline
		Filter          & FWHM ($\arcsec$)   \\
		\hline
F435W	&	0.107	\\
F606W	&	0.111	\\
F775W	&	0.107	\\
F814W	&	0.118	\\
F850LP	&	0.108	\\
F105W	&	0.181	\\
F125W	&	0.179	\\
F140W	&	0.181	\\
F160W	&	0.184	\\

		\hline
	\end{tabular}
\end{table}

\begin{figure*} 
\centering
\includegraphics[width=0.8\hsize]{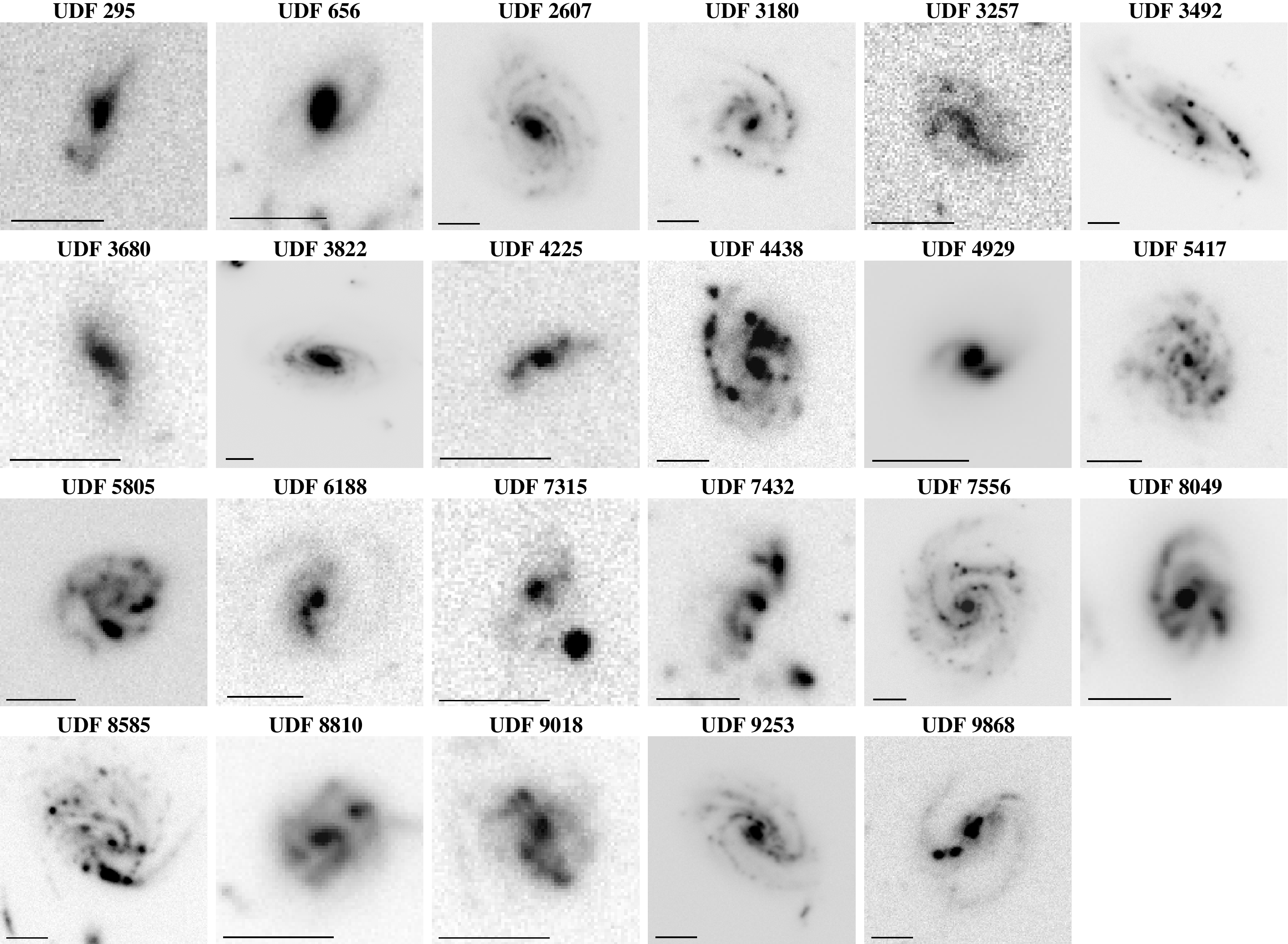}
\caption[fig2]{Combined 30 mas images of spiral galaxy candidates.
The horizontal line in the lower left corner of each frame represents 1\arcsec.
The display is in linear scale. North is up, east to the left.
}
~\label{fig2}
\end{figure*}
\begin{figure} 
\centering
\includegraphics[width=1.0\hsize]{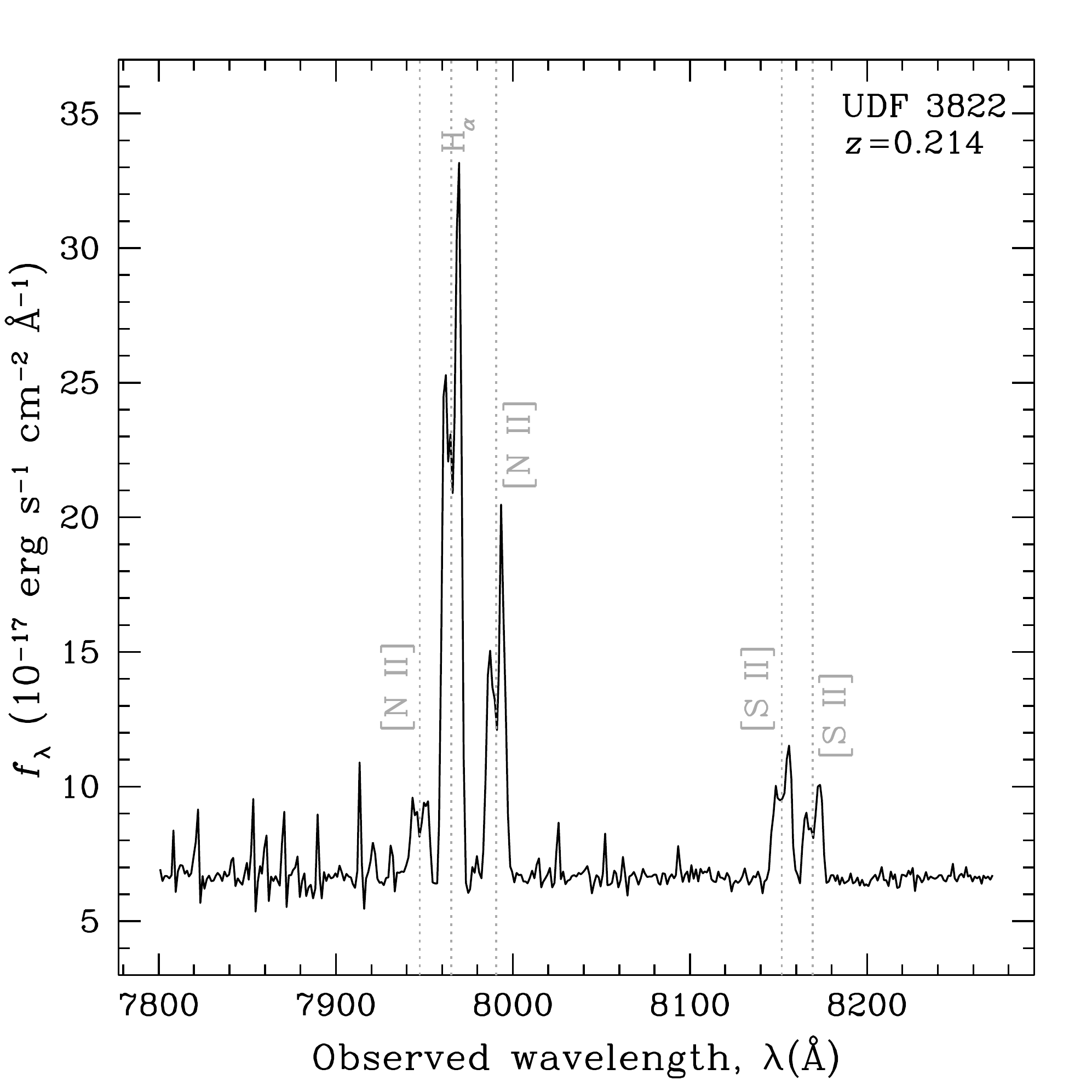}
\caption[fig3]{Zoom-in on the MUSE spectrum of the galaxy UDF~3822, integrated
over all spatial pixels.
Some emission lines are labelled.
}
~\label{fig3}
\end{figure}

We combined the 30 mas images into a single frame (thus maximizing the signal-to-noise ratio, S/N).
Next, we split the image into $12\arcsec\times12\arcsec$ boxes covering the entire image,
which measures approximately $\sim200\arcsec\times200\arcsec$.
We focused on selecting grand design, non-barred (SA) or weakly barred (SAB) spiral galaxies.\footnote{
Spirals arms in barred galaxies (SB) may have an origin other than density waves,
e.g., chaotic orbits guided by invariant `manifolds'~\citep[see, e.g.,][]{rom06,pat06,mart12,atha12,con22}.
This theory also predicts a constant spiral pattern speed, although
with lifetimes shorter than those anticipated by density wave theory.}
The initial candidate sample consisted of 23 objects (see Figure~\ref{fig2}).
We verified that each candidate has a rotating disc.
To achieve this goal we used data from the Multi-Unit Spectroscopic Explorer~\citep[MUSE,][]{bac10}
integral-field spectrograph on the Very Large Telescope (VLT).
We retrieved from the European Southern Observatory (ESO) Science Archive data
belonging to the MUSE Hubble Ultra Deep Survey~\citep[HUDF;][]{bac17}, and the MUSE Extremely Deep Field~\citep{bac21}.
The MUSE HUDF data cubes have a spatial sampling (pixel scale) of $\sim0\farcs2$,
and a spectral sampling of 1.25~\AA~over the wavelength range 4750~\AA~$-$~9350~\AA.
The PSF FWHM of the MUSE data is $\sim0\farcs71$ at 4750~\AA, and $\sim0\farcs57$ at 9350~\AA, 
owing to the ESO adaptive optics methods.
We identified each one of our objects in the MUSE cubes, subtracted the local sky,
and obtained moment-1 (velocity) maps by using:

\begin{equation}
      v = c\left(\frac{\int_{\lambda_{0}}^{\lambda_{1}} \lambda I_{\lambda} {\rm d}\lambda}{(1+z)\lambda_{\rm rest} \int_{\lambda_{0}}^{\lambda_{1}} I_{\lambda} {\rm d}\lambda} - 1\right),
\end{equation}

\noindent where $v$ is the observed velocity, $c$ is the speed of light,
$\lambda$ is the observed wavelength, $I_{\lambda}$ is the intensity at wavelength $\lambda$,
$\lambda_{0}-\lambda_{1}$ is the observed wavelength range of a certain emission line,
$z$ is the redshift of the galaxy,
and $\lambda_{\rm rest}$ is the rest wavelength of the emission line.
The brightest emission lines for various HUDF objects are given in~\citet{ina17}.
For this work we use emission lines devoid of residual sky lines.
Depending on the object, the adopted emission lines were one among
[\ion{O}{ii}] $\lambda$3726, [\ion{O}{ii}] $\lambda$3729,\footnote{The [\ion{O}{ii}] 3726-3729~\AA~doublet was treated as two separate lines.}
H\,$\beta$, and H\,$\alpha$.
In Figure~\ref{fig3} we show the spectrum near the H\,$\alpha$ line used to derive the moment-1 map
of UDF~3822. Similar spectra were obtained for most of the objects in our sample.
From this analysis we were able to confirm disc rotation in 16 galaxies.
There are no data for UDF~295 in the MUSE HUDF.
For the rest of the objects the spatial resolution of the MUSE cubes is poor, hence we could not confirm or disprove disc rotation.
These objects are UDF~295, UDF~3257, UDF~3680, UDF~4225, UDF~6188, UDF~7315, and UDF~7432.
We retrieved data for them from the Atacama Large Millimeter/submillimeter Array (ALMA) Science Archive.\footnote{
Project codes: ADS/JAO.ALMA\#2012.1.00173.S, ADS/JAO.ALMA\#2015.1.00098.S, ADS/JAO.ALMA\#2015.1.00543.S,
ADS/JAO.ALMA\#2016.1.00324.L, and ADS/JAO.ALMA\#2017.1.00138.S.
ALMA is a partnership of ESO (representing its member states), NSF (USA) and NINS
(Japan), together with NRC (Canada), MOST and ASIAA (Taiwan), and KASI (Republic
of Korea), in cooperation with the Republic of Chile. The Joint ALMA Observatory is
operated by ESO, AUI/NRAO and NAOJ. The National Radio Astronomy Observatory is a facility of the National Science Foun-
dation operated under cooperative agreement by Associated Universities, Inc.}
Unfortunately, we likewise could not find evidence of disc rotation,
due to either a synthesised beam larger than the angular size of the objects,
or to lack of bright emission lines. We exclude these galaxies from further analysis. 

\begin{figure*} 
\centering
\includegraphics[width=1.0\hsize]{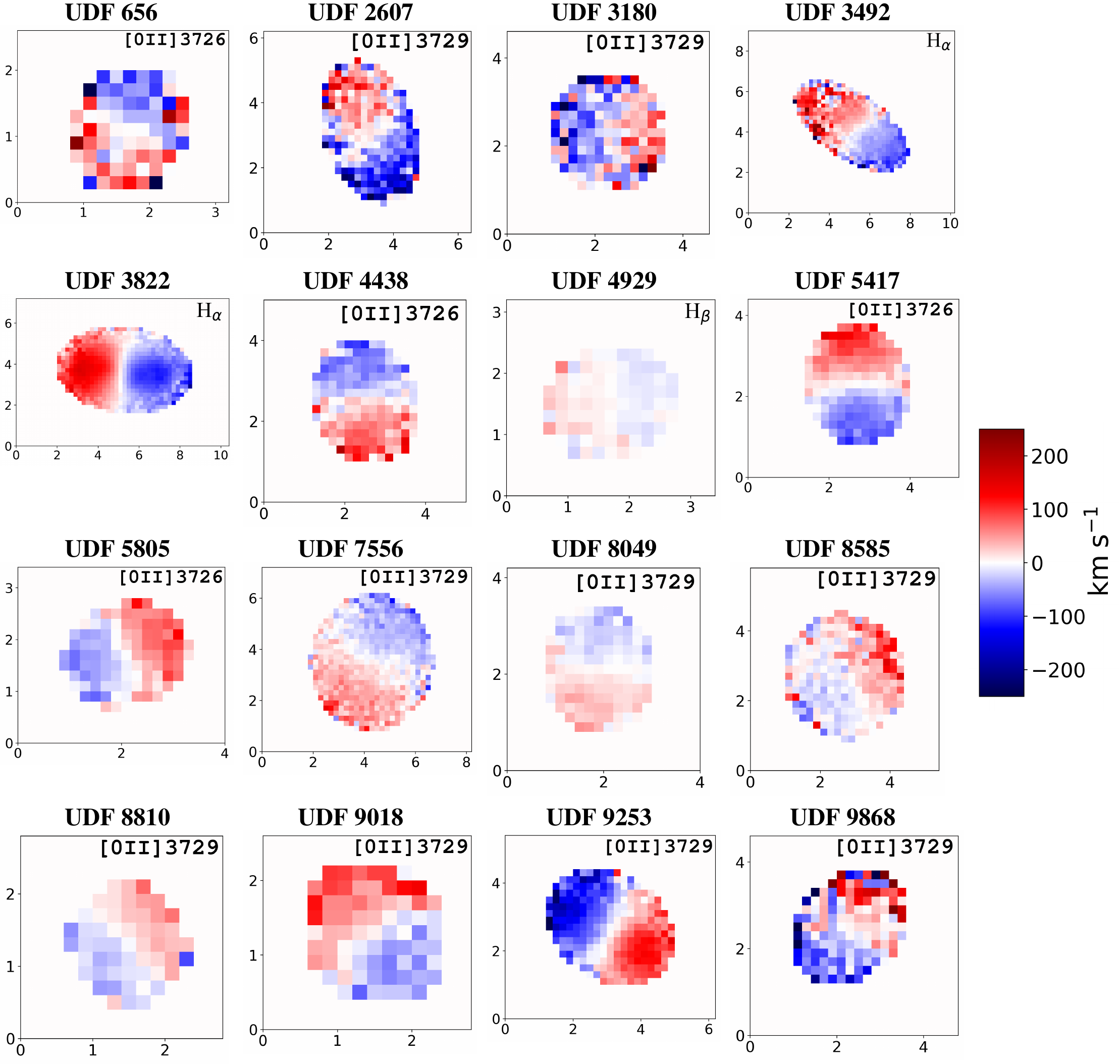}
\caption[fig4]{Moment-1 (velocity) maps of the galaxy sample (see Table~\ref{tbl-2}). Axis units are in arcseconds (\arcsec).
}
~\label{fig4}
\end{figure*}

\begin{figure} 
\centering
\includegraphics[width=\columnwidth]{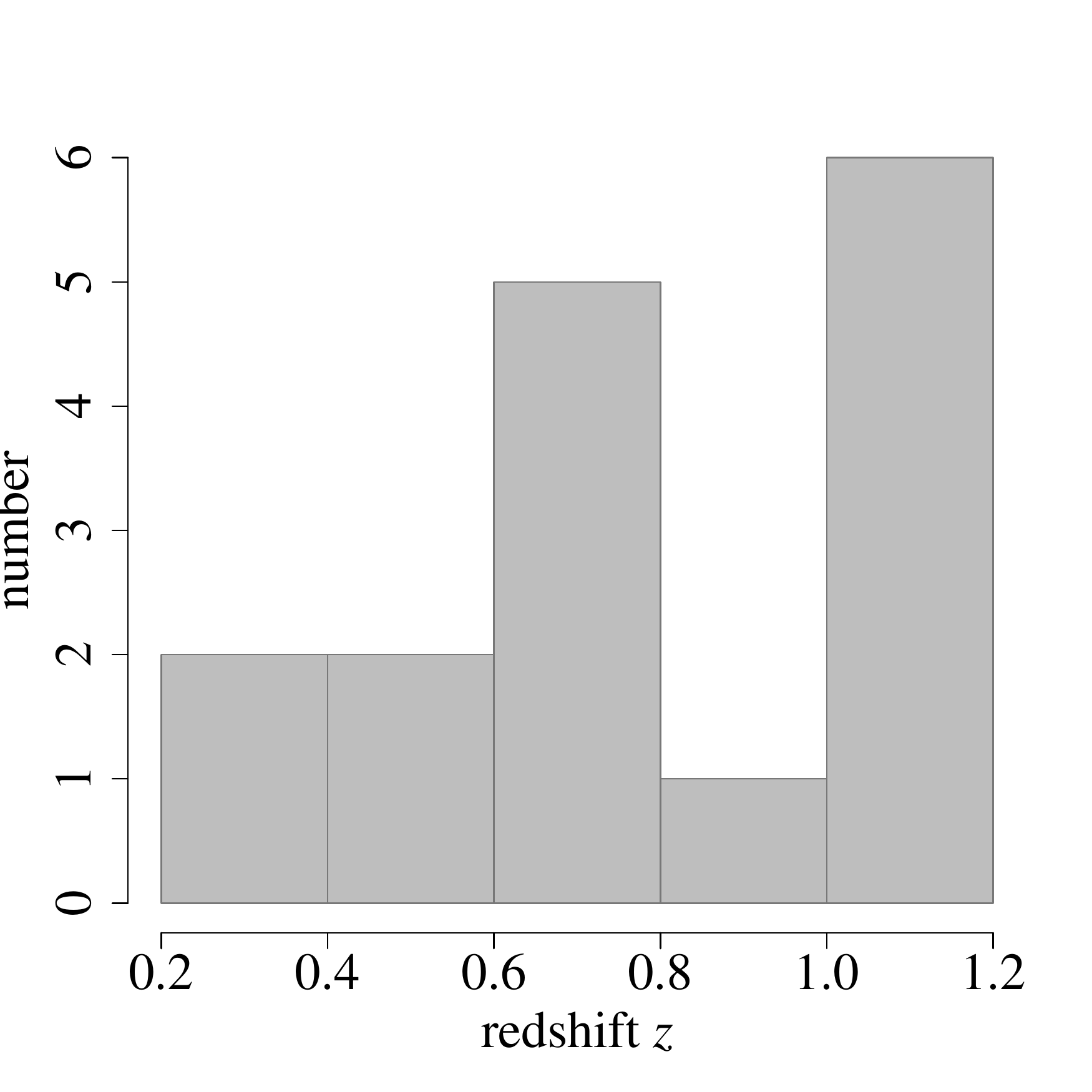}
\caption[fig5]{Histogram of the spectroscopic redshifts of our galaxy sample.
}
~\label{fig5}
\end{figure}

\begin{figure} 
\centering
\includegraphics[width=\columnwidth]{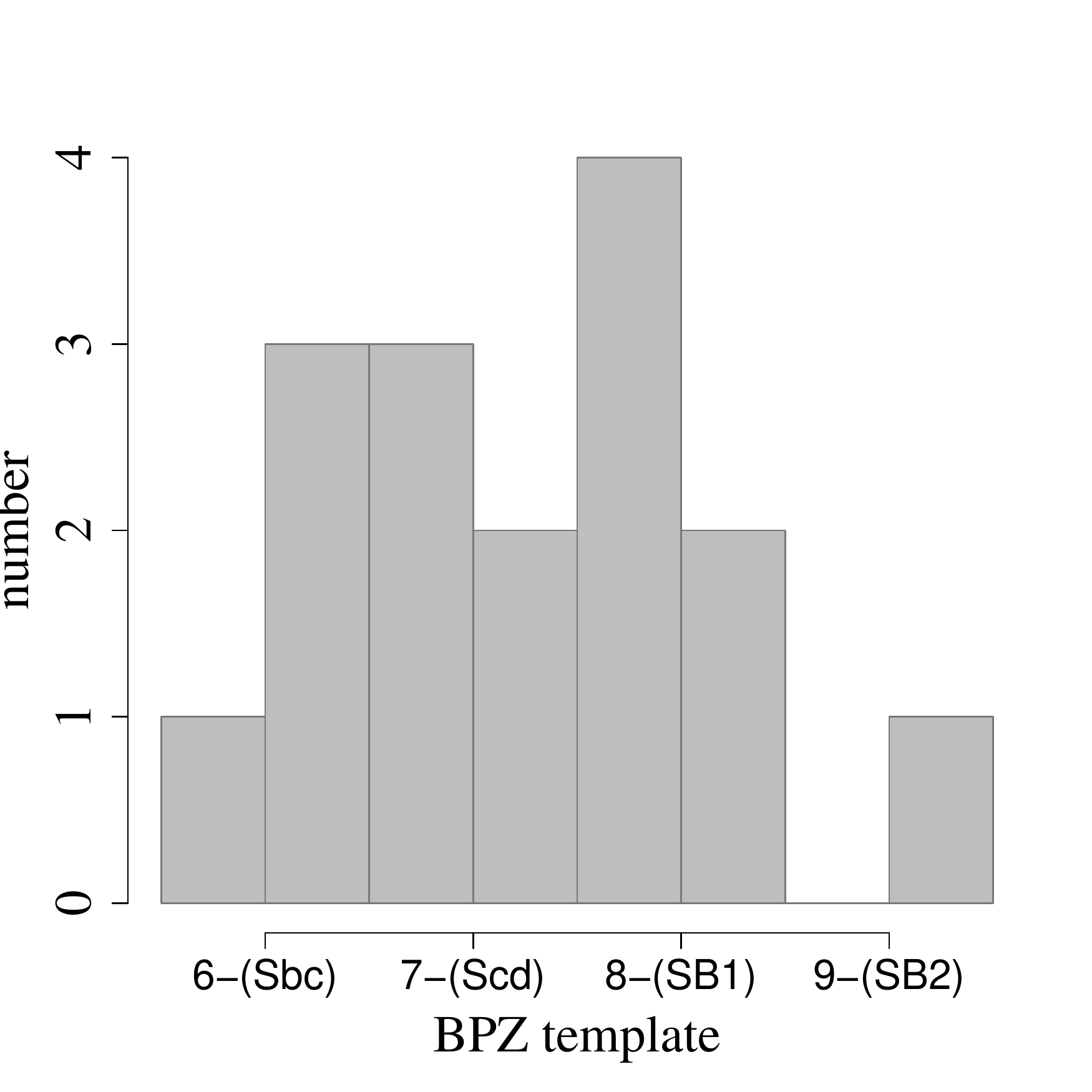}
\caption[fig6]{Histogram of SED templates fitted to the sample~\citep{rafel15}.
Spirals are designated Sbc and Scd (template numbers 6 and 7), and starbursts are designated SB1 and SB2 (template numbers 8 and 9).
}
~\label{fig6}
\end{figure}

Our final sample consists of 16 objects that are listed in Table~\ref{tbl-2}.
Moment-1 maps are shown in Figure~\ref{fig4}.
Spectroscopic redshifts were taken from~\citet{ina17}; a histogram
of the redshift distribution is shown in Figure~\ref{fig5}.
Also in Table~\ref{tbl-2}, we list in column 6 the spectral energy
distribution (SED) templates fitted to obtain the photometric redshift of each object, 
as described in~\citet[][; see also~\citet{ben00},~\citet{coe06}]{rafel15}.
A histogram of the template types is shown in Figure~\ref{fig6}.
The SED templates include spiral galaxies, with Hubble types
Sbc and Scd (template numbers 6 and 7, respectively);
starburst galaxies (SB, template numbers 8-11);
and lenticular galaxies (template number 5).
Templates with intermediate numbers are created by interpolation.
The best fit SED template may imply a preliminary galaxy type classification.

We also corroborate that our objects have active star formation in their discs.
For this purpose we use the Code Investigating GALaxy Emission~\citep[{\tt{CIGALE}},][]{bur05,noll09,boq19};
we employ the aperture photometry measurements of~\citet{rafel15}, 
and include the ultra-violet filters when available.
The instantaneous star formation rate (SFR), $\Psi$, is given in
Table~\ref{tbl-3},\footnote{See Appendix~\ref{appA} for the input configuration adopted in \tt{CIGALE}.}
together with the average SFR over the last 10 Myr,  $\langle\Psi\rangle^{\rm 10~Myr}$,
and the average SFR over the last 100 Myr, $\langle\Psi\rangle^{\rm 100~Myr}$.
All of our objects are actively forming stars, i.e., they have an SFR, $\Psi>1$ M$_{\sun}$ yr$^{-1}$.

\begin{table*}
	\centering
	\caption{Spiral galaxy sample.}
	\label{tbl-2}
	\begin{tabular}{cccccc} 
		\hline
		UDF-ID          & R.A. (J2000, deg)     & Decl. (J2000, deg) & UVUDF-ID        & Spectroscopic redshift $z$  & SED Template \\
		~\citep{coe06}  & (this work)           & (this work)        &~\citep{rafel15} &~\citep{ina17}   &~\citep{rafel15} \\
		\hline


656   &  53.164455  &  -27.815337  &  830    &  1.096718	&  6.2  \\
2607  &  53.180240  &  -27.798929  &  21364  &  0.667092	&  6.2  \\
3180  &  53.157823  &  -27.797522  &  3103   &  0.767559	&  8.5  \\
3492  &  53.187823  &  -27.794047  &  51705  &  0.345790	&  8.5  \\
3822  &  53.186947  &  -27.790995  &  22719  &  0.213732	&  6.6  \\
4438  &  53.137642  &  -27.791988  &  52744  &  0.997884	&  7.8  \\
4929  &  53.187952  &  -27.789999  &  22718  &  0.435454	&  6.3  \\
5417  &  53.166166  &  -27.787515  &  22410  &  1.095512	&  7.3  \\
5805  &  53.192043  &  -27.787163  &  34879  &  1.095373	&  7.9  \\
7556  &  53.169922  &  -27.771048  &  24587  &  0.622005	&  6.9  \\
8049  &  53.162345  &  -27.775059  &  24348  &  0.419313	&  7.3  \\
8585  &  53.147860  &  -27.774034  &  54454  &  1.087423	&  7.9  \\
8810  &  53.155280  &  -27.769540  &  24453  &  0.736333	&  7.7  \\
9018  &  53.147061  &  -27.778419  &  9759   &  1.086319	&  9.2  \\
9253  &  53.178387  &  -27.768230  &  24420  &  0.668620	&  6.7  \\
9868  &  53.163590  &  -27.758934  &  37143  &  1.094886	&  5.7  \\

		\hline
	\end{tabular}
\end{table*}


\begin{table}
	\caption{Star formation rates, $\Psi$.}
	\label{tbl-3}
	\begin{tabular}{cccc} 
		\hline		
                UDF-ID  & $\Psi$                  & $\langle\Psi\rangle^{\rm 10~Myr}$  & $\langle\Psi\rangle^{\rm 100~Myr}$ \\
                     ~  & (M$_{\sun}$ yr$^{-1}$)  & (M$_{\sun}$ yr$^{-1}$)             & (M$_{\sun}$ yr$^{-1}$)             \\
                \hline

656	&	7.5	$\pm$	3.0	&	7.5	$\pm$	3.1	&	7.9	$\pm$	4.1	\\
2607	&	19.7	$\pm$	15.6	&	20.1	$\pm$	16.4	&	27.1	$\pm$	32.0	\\
3180	&	4.5	$\pm$	0.2	&	4.8	$\pm$	0.2	&	9.6	$\pm$	0.5	\\
3492	&	2.9	$\pm$	1.7	&	3.0	$\pm$	1.8	&	6.1	$\pm$	3.3	\\
3822	&	7.2	$\pm$	6.6	&	7.5	$\pm$	7.0	&	13.1	$\pm$	14.6	\\
4438	&	10.6	$\pm$	1.8	&	11.1	$\pm$	1.8	&	20.2	$\pm$	2.4	\\
4929	&	4.5	$\pm$	1.0	&	4.5	$\pm$	1.0	&	4.6	$\pm$	1.0	\\
5417	&	21.8	$\pm$	7.8	&	22.2	$\pm$	7.8	&	28.8	$\pm$	10.8	\\
5805	&	7.6	$\pm$	2.9	&	7.9	$\pm$	3.0	&	13.7	$\pm$	5.2	\\
7556	&	16.7	$\pm$	4.5	&	16.9	$\pm$	4.4	&	19.8	$\pm$	5.0	\\
8049	&	3.6	$\pm$	0.9	&	3.8	$\pm$	0.9	&	6.6	$\pm$	1.5	\\
8585	&	32.1	$\pm$	5.7	&	33.6	$\pm$	5.7	&	60.4	$\pm$	6.7	\\
8810	&	4.2	$\pm$	0.9	&	4.4	$\pm$	0.9	&	7.8	$\pm$	1.3	\\
9018	&	3.7	$\pm$	0.2	&	3.9	$\pm$	0.2	&	7.8	$\pm$	0.4	\\
9253	&	16.1	$\pm$	4.2	&	16.2	$\pm$	4.2	&	17.8	$\pm$	4.8	\\
9868	&	17.5	$\pm$	15.9	&	17.6	$\pm$	15.9	&	18.3	$\pm$	16.8	\\

		\hline																																																				
	\end{tabular}
\end{table}


\section{Analysis}~\label{analysis}

For the sample galaxies we obtain the projection parameters, i.e.,
the inclination and position angles (I.A and P.A., respectively, see Table~\ref{tbl-4}),
by fitting ellipses to their outer isophotes in the combined 30 mas images.
After deprojecting the images, we proceeded to measure the pitch angles, $P$,
of the spiral arms in each individual 30 and 60 mas image,
by way of Fourier transform techniques.
We first transform the images from Cartesian ($x,y$)
to logarithmic polar coordinates $\ln{R}$, $\theta$; $\ln{R}\equiv{u}$;
we then compute the two-dimensional Fourier transform~\citep[e.g.,][]{pue92,sar94}:

\begin{equation}~\label{eq2DFT}
A(m,p)=\frac{\int_{u_{\mathrm{min}}}^{u_{\mathrm{max}}} \int_{-\pi}^{\pi}
I(u,\theta)e^{-i(m\theta+pu)}d\theta du}{\int_{u_{\mathrm{min}}}^{u_{\mathrm{max}}} \int_{-\pi}^{\pi}
I(u,\theta)d\theta du},
\end{equation}

\noindent where $I(u,\theta)$ is the radiation intensity of the pixel with coordinates ($u,\theta$).
The values $u_{\mathrm{min}},u_{\mathrm{max}}$ delimit the radial region
of the spiral arms given by $\Delta{R}=(R_{\mathrm{min}}-R_{\mathrm{max}})$.
In order to avoid potential biases we use the same $\Delta{R}$
at all wavelengths/passbands for the same object~\citep[see also][]{{yu18}}.
The adopted $\Delta{R}$ values are given in Table~\ref{tbl-4}
in units of arcseconds and in kpc~\citep[taking into account cosmological effects,][]{wri06,ben14};
they were chosen by visually inspecting the images
in order to include the spiral arms and avoid the central bulge.
The cosmological surface brightness dimming, $\sim~(1+z)^{-4}$, may affect our $P$ measurements,
in the sense that higher-redshift galaxies may be measurable only in their inner regions.
This affects mostly diffuse, rather than compact, objects~\citep[e.g.,][]{cal14}.
However, even if the cosmological surface brightness dimming affects the observed spiral loci
of our galaxies, any differences of $P$ with wavelength will still be detected.

\begin{figure} 
\centering
\includegraphics[width=\columnwidth]{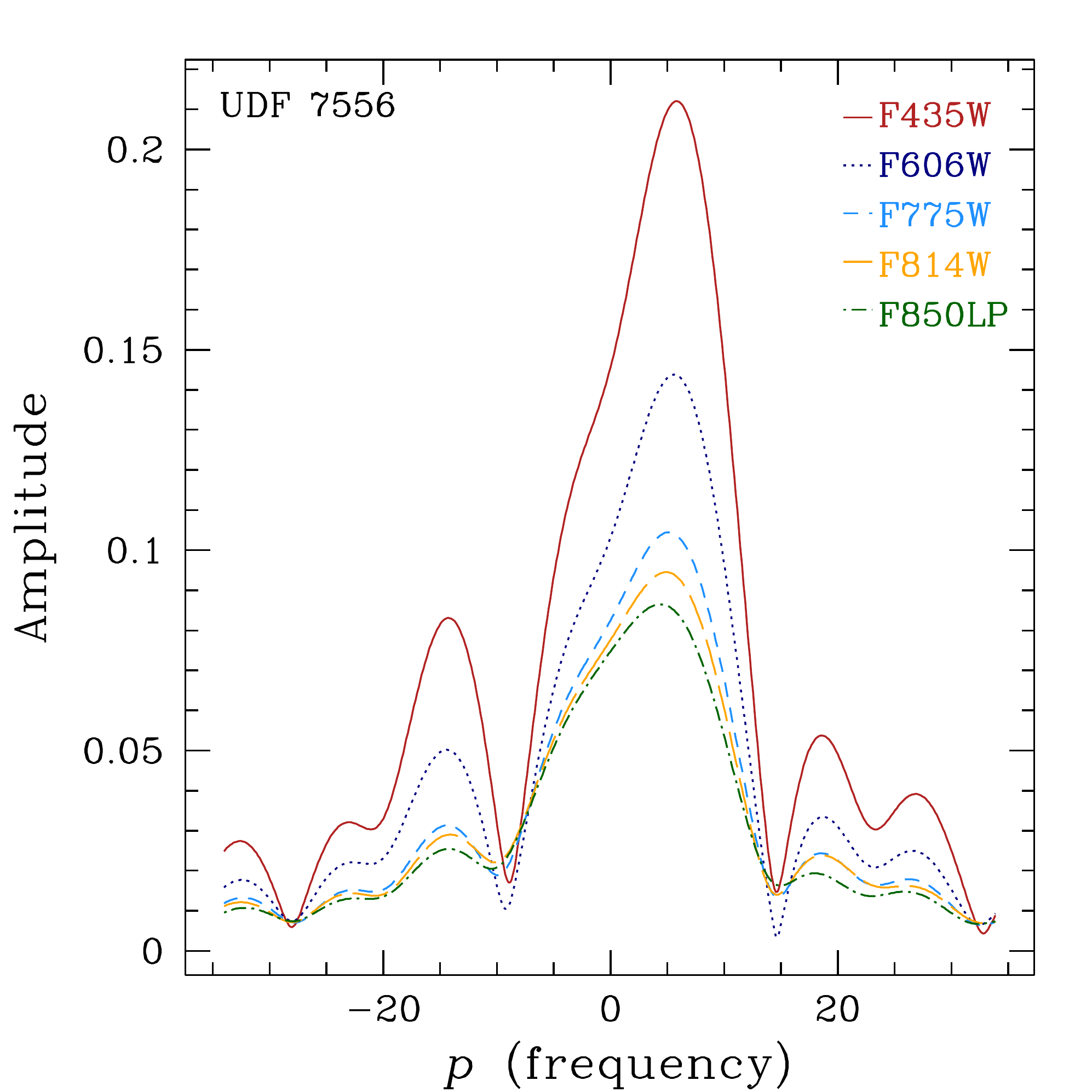}
\caption[fig7]{$A(m=2,p)$ spectra for UDF~7556, computed for the 30 mas images in filters F435W, F606W, F775W, F814W, and F850LP.
The maximum amplitude determines the value of $p_{\mathrm{max}}$ for each filter (see equation~\ref{eqP}).
}
~\label{fig7}
\end{figure}

From the output $A(m,p)$ matrix we derive the pitch angle with:
\begin{equation}~\label{eqP}
 \tan{P} = -m/p_{\mathrm{max}},
\end{equation}
\noindent where $m=0,1,2,3\dots$ All our objects have a bisymmetric spiral, hence
we adopt $m=2$ for all the calculations.
The values of $p_{\mathrm{max}}$ at each wavelength and for each object are
found by computing the absolute maximum of the $A(m=2,p)$ spectra.
In Figure~\ref{fig7} we show the $A(m=2,p)$ spectra for UDF~7556, at different wavelengths.

Star-forming clumps are less abundant in nearby galaxies
than at higher redshifts~\citep[e.g.,][]{add22,sat23}.
Since these clumps are assumed to be related to star formation, no masks were applied
to them. Also, the Fourier transform technique described above separates the signals
of the different components contributing to the luminosity in the image. For this reason, the $P$ measurements are not expected
to show significant differences in the presence of massive star-forming clumps, unless
their presence effectively suppresses the signal from the spiral arms.
This is not the case for the galaxies in our sample (see Figure~\ref{fig2}). 

\subsection{The effect of angular size for distant galaxies}~\label{sect_ang_size}

\begin{figure} 
\centering
\includegraphics[width=\columnwidth]{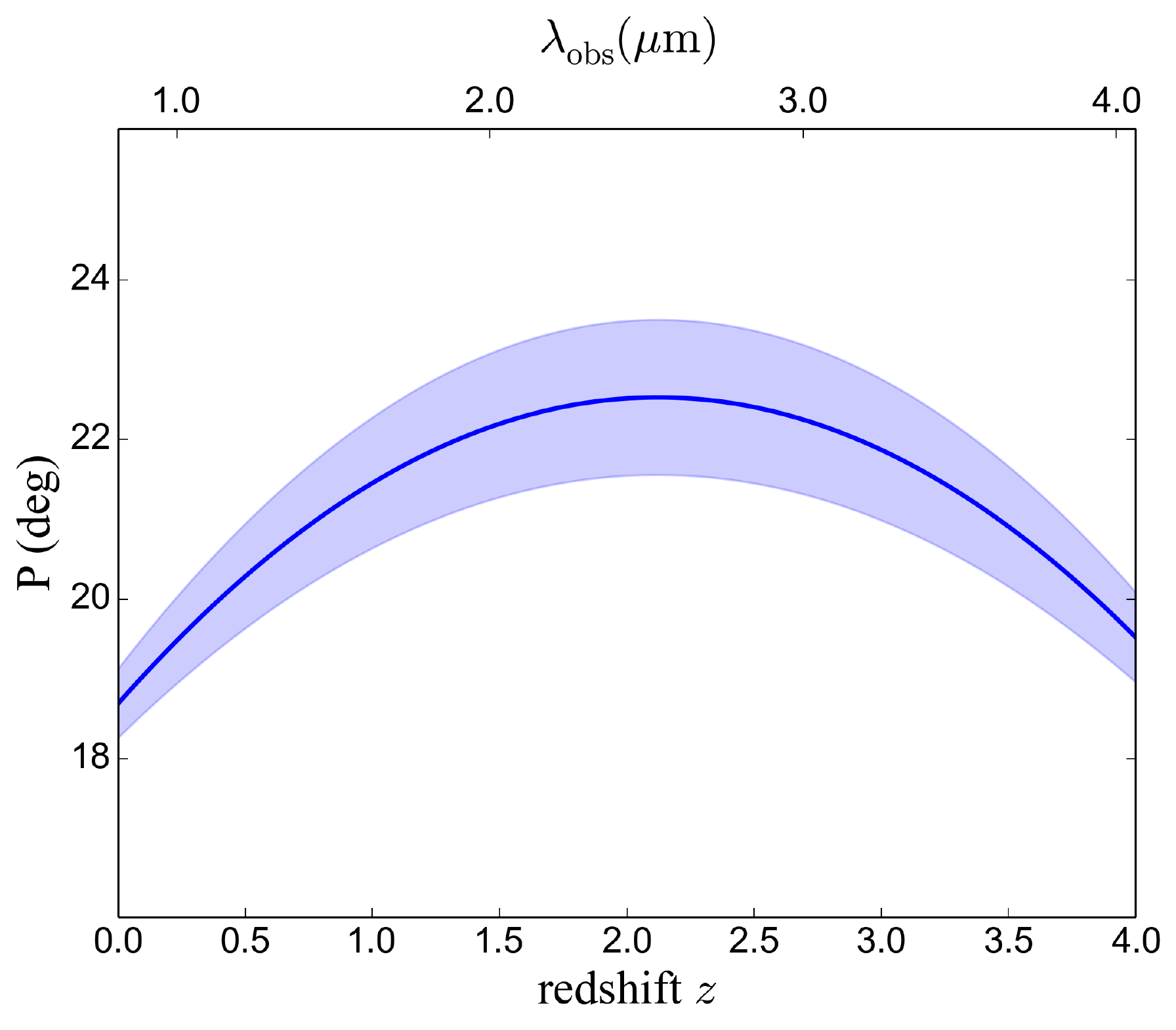}
\caption[fig8]{Pitch angles of M51 (NGC~5194), as a function of simulated redshift $z$. 
The $z=0$ data corresponds to the {\sl HST} F814W image.
{\it Solid blue line:} mean value; {\it shaded region:} $\pm\ 1\ \sigma$ error.
The upper $x$ axis denotes the observed wavelength, $\lambda_{\rm obs} = \lambda_{\rm rest} (1+z)$.
}
~\label{fig8}
\end{figure}

\begin{figure} 
\centering
\includegraphics[width=\columnwidth]{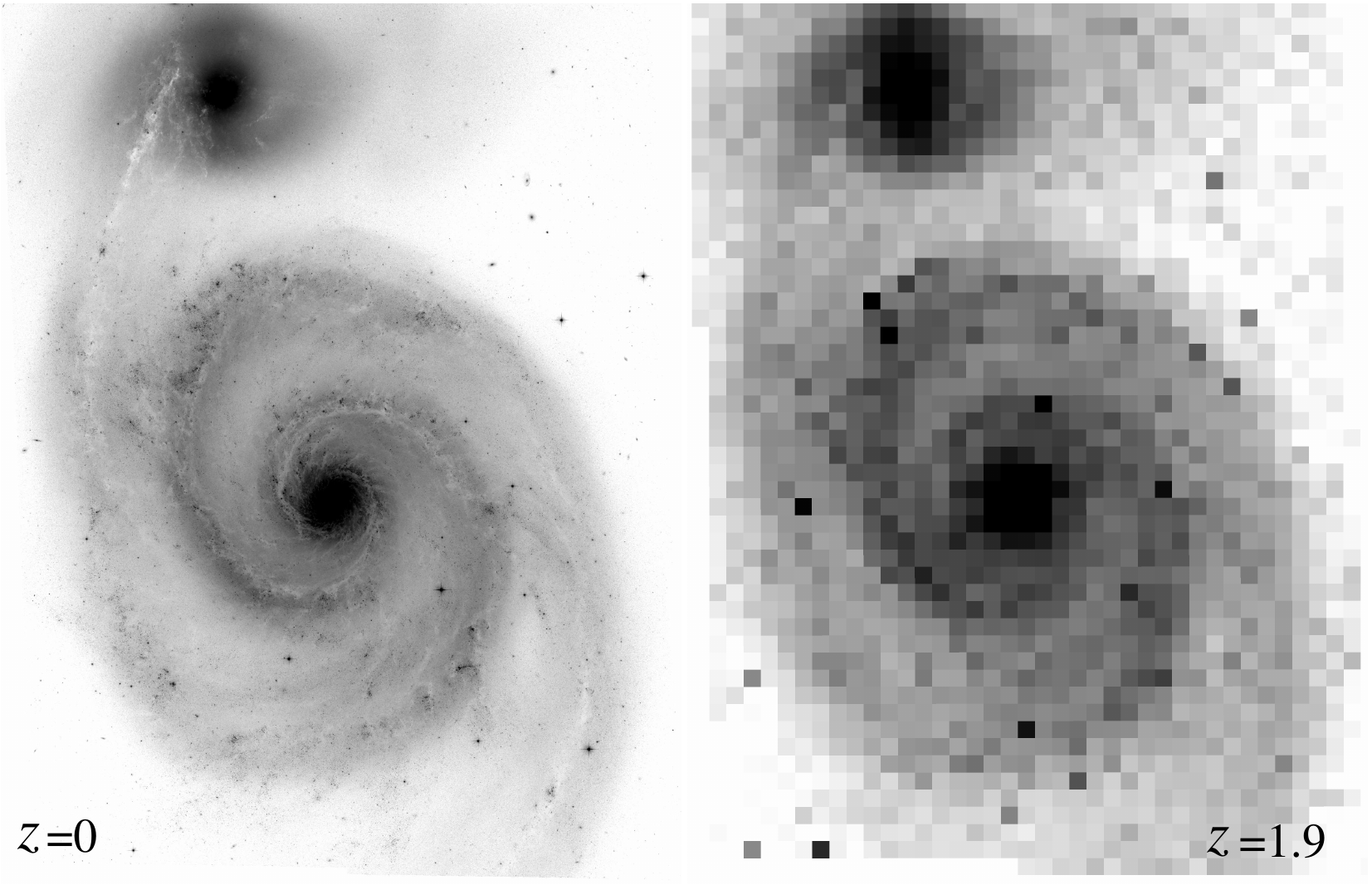}
\caption[fig9]{{\it Left:} Mosaic of the spiral galaxy M51 in the {\sl HST} F814W filter at redshift $z=0$. 
{\it Right:} Simulated image of M51 at redshift $z=1.9$, $\lambda_{\rm obs}\sim2.4\mu{\rm m}$,
obtained by assuming the same pixel scale as the F814W image.
}
~\label{fig9}
\end{figure}

In this section we explore the effect that angular size has
on our measurements of pitch angles for disc galaxies at redshifts $z>0$~\citep[see also][]{blo01}.
For this purpose we simulate the appearance of the grand design spiral galaxy M51 (NGC~5194) at different redshifts.
We use the {\sl HST} F814W-band mosaic of M51 obtained from the Hubble Legacy Archive.
This image was secured with the ACS WFC, whose pixels are 0$\farcs$05 on the side.
We assume a distance of 8.58 ± 0.10 Mpc~\citep[$z\sim0$,][]{mcq16}, an inclination angle of 20\degr,
and a position angle of 172\degr~\citep{ler08}.
At the assumed distance of M51, 0$\farcs$05 is $\sim$ 2 pc.
We use the cosmological calculator of~\citet{wri06} to derive the angular size of M51 at
different redshifts. The results of this simulation, at fixed pixel scale, are shown in Figure~\ref{fig8}.
At $z\sim0$, the pitch angle is $P\sim19\degr$~\citep{pue14}.
At redshift $z\sim2$, one ACS WFC pixel is $\sim$ 400 pc, 
the observed wavelength is $\lambda_{\rm obs}\sim2.4\mu{\rm m}$ and the mean $P$ value increases to $P\sim22\degr$,
with a larger $1\sigma$ error. For redshifts higher than $z\gtrsim2$, the $P$ value decreases again
due to the standard cosmological model, which predicts an angular diameter turnover point~\citep[e.g.,][]{mel18}.\footnote{
This is a cosmological effect in the apparent angular diameter due to the expansion of
the universe and the finite speed of light.}
The higher $P$ values at intermediate redshifts are due to the `pixellation' of the spiral arms: as we increase
the redshift of M51, and keep the pixel scale fixed, fewer pixels
trace the spiral arms in the image. In Figure~\ref{fig9} we show the M51 image at redshift $z=0$ (left frame)
and at $z=1.9$ (right frame). The simulated $z=1.9$ galaxy
is $\sim200$ times smaller in angular size and hence encompasses many fewer pixels.
From this exercise we can conclude that the pitch angles for galaxies at $0<z<2$
may suffer from `pixellation', and result in artificially higher pitch angles and errors.
These effect is more deleterious for images with coarser plate scales. For instance,
if we use the Sloan Digital Sky Survey (SDSS) $i$-band mosaic of M51~\citep[plate scale = 0$\farcs$396~pixel$^{-1}$,][]{anh14,ala15},
we find that the pitch angle increases by a factor of $\sim1.5$ times,
and the $1\sigma$ error grows by a factor of $\sim6$ in the redshift range $0<z<0.16$.\footnote{
For $z\gtrsim0.16$, the spiral arms of M51 are no longer distinguishable for images with the same pixel scale as the SDSS mosaic.}
It is important to mention that this effect affects the $P$ measurements at every $\lambda$ in a similar manner.
That being said, at a fixed redshift, the difference in the measured pitch angles at different wavelengths will still
be detected if present in real galaxies.

\subsection{Differences in the PSF.}

\begin{figure} 
\centering
\includegraphics[width=\columnwidth]{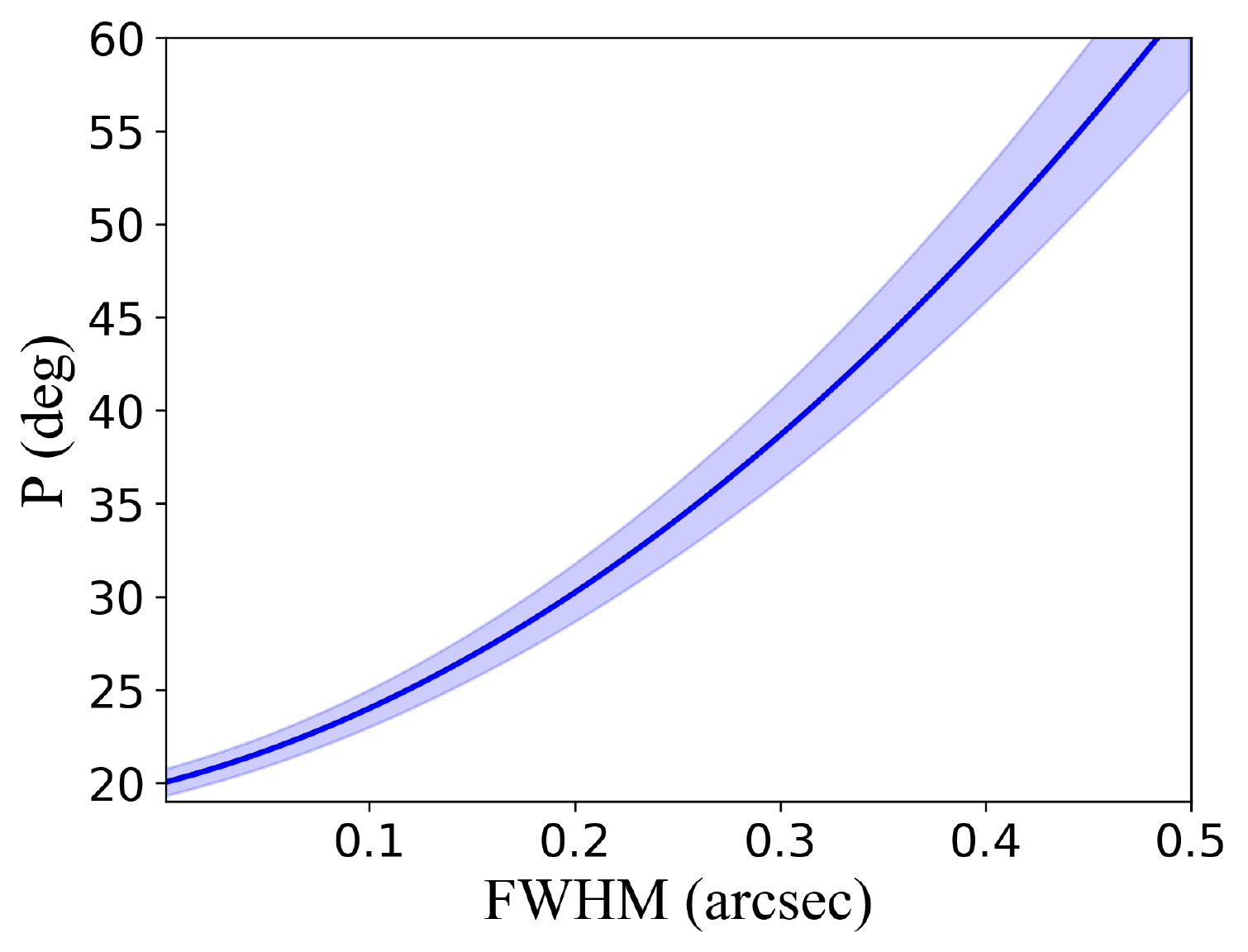}
\caption[fig10]{Pitch angles of the simulated M51 image, at redshift $z=1.4$, after convolving the
image with Gaussian kernels with different FWHMs (i.e., different PSF).
}
~\label{fig10}
\end{figure}

For distant objects, which are imaged by considerably fewer pixels than nearby objects,
a difference in the PSF for different filters may produce faulty pitch angle measurements.
In the case of the 30 mas XDF data, the PSF FWHM is practically the same at all wavelengths~\citep[$\sim0\farcs1$,][]{win11},
and no corrections are needed.
This is not the case for the 60 mas data (see Table~\ref{tbl-1}), where the four
NIR images have a PSF FWHM $\sim1.6$ times larger than the optical images.
In order to test the impact of a different PSF on the observations of distant spiral galaxies,
we use the simulated M51 image at redshift $z\sim1.4$, produced in Section~\ref{sect_ang_size}
from the Hubble Legacy data. We convolve this image with a Gaussian
kernel with different values of $\sigma$.
The result of this exercise is displayed in Figure~\ref{fig10}.
As shown in the plot, a larger FWHM results in a higher $P$ value.
This effect may bias our measurements of the pitch angles.
With the purpose of minimising this effect on the 60 mas data,
and since we are interested in pitch angle relative differences, rather than in their absolute values,
we correct the optical images (filters F435W, F606W, F775W, F814W, and F850LP)
to match the NIR PSF FWHM (filters F105W, F125W, F140W, and F160W).
We assume that the PSF matching can be achieved through a
convolution with a Gaussian kernel~\citep[see also][]{ani11}, i.e., we posit
that the output PSF FWHM is the result of the convolution of two Gaussian functions,
$\sigma_{f*g}=\sqrt{\sigma^2_f+\sigma^2_g}$, where $\sigma_f$ and $\sigma_g$ are the 
standard deviations of the two original Gaussians, and the result of the convolution is also Gaussian.
The 60 mas data were corrected before analysis.

We also tested the effect of a different PSF on the nearby galaxy M51 by
using far and near ultraviolet ($FUV$ and $NUV$, respectively) imaging from the {\it GALEX}
Ultraviolet Atlas of Nearby Galaxies~\citep{gil07}, and
$u$-, $g$-, $r$-, $i$- and $z$-band optical data from SDSS~\citep{anh14, ala15}.
We registered the SDSS images to the spatial resolution
of the $NUV$ image (which has a plate scale of $1.5\arcsec$ pixel$^{-1}$).
The GALEX images have a PSF with a FWHM a factor of $\sim$4 larger than SDSS frames.
In order to get a common PSF for all our data, we use the 2017 version of the convolution kernels of~\citet{ani11}.
We found that the pitch angles obtained from the PSF-corrected optical images are slightly higher ($\sim 0.4\degr$) 
than the ones without it; however, this difference is similar to the computation error of the method.
Hence, we can conclude that, for nearby galaxies, a different PSF does not produce, in practice,
a variation on the pitch angle measurements.

\section{Results and discussion}~\label{results}

\begin{figure} 
\centering
\includegraphics[width=\columnwidth]{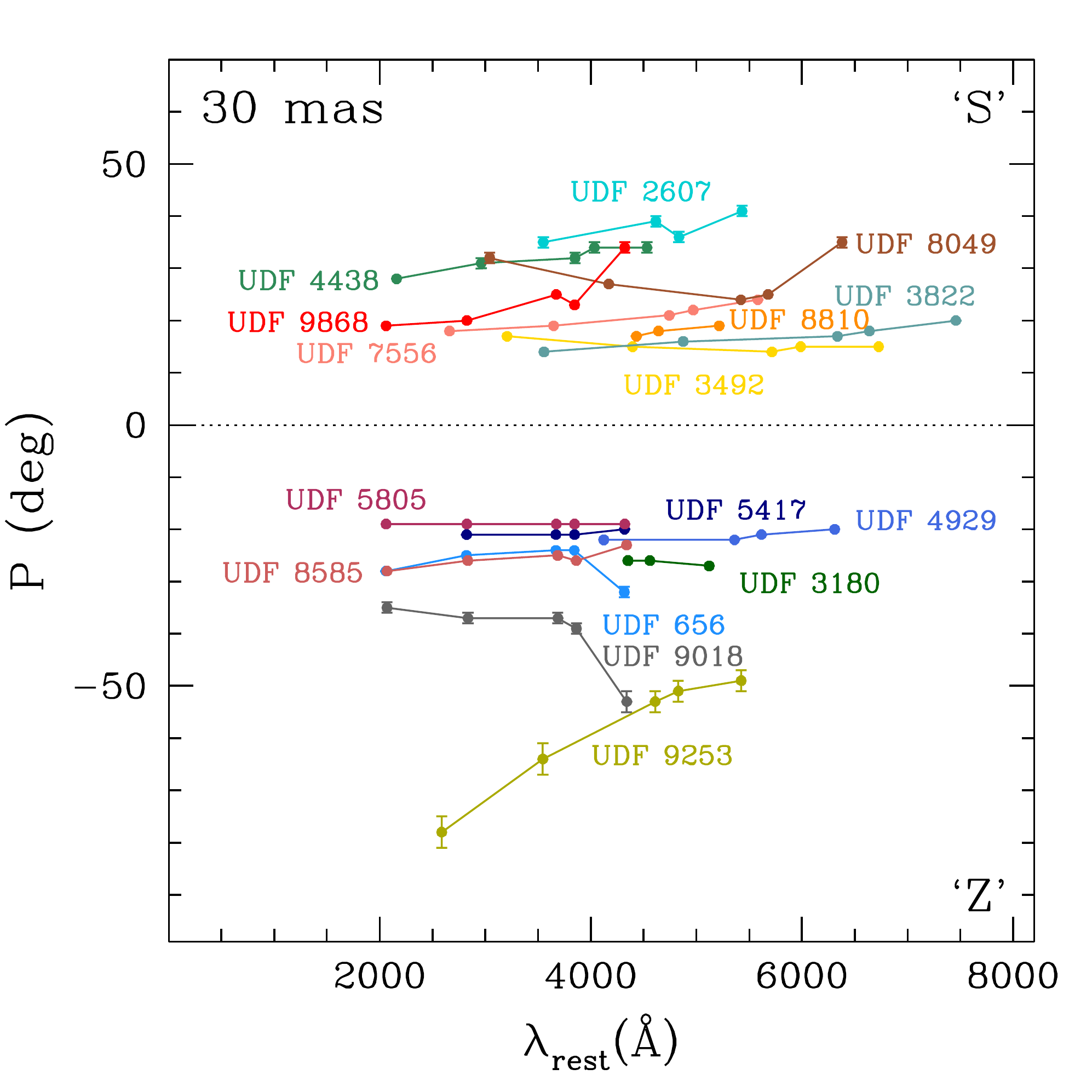}
\caption[fig11]{Measured pitch angles, $P(\deg)$, as a function of rest
wavelength, $\lambda_{\rm rest}$(\AA), for the 30 mas XDF images in filters
F435W, F606W, F775W, F814W, and F850LP.
`S' or `Z' indicate the galaxies on-the-sky view. 
}
~\label{fig11}
\end{figure}

\begin{figure} 
\centering
\includegraphics[width=\columnwidth]{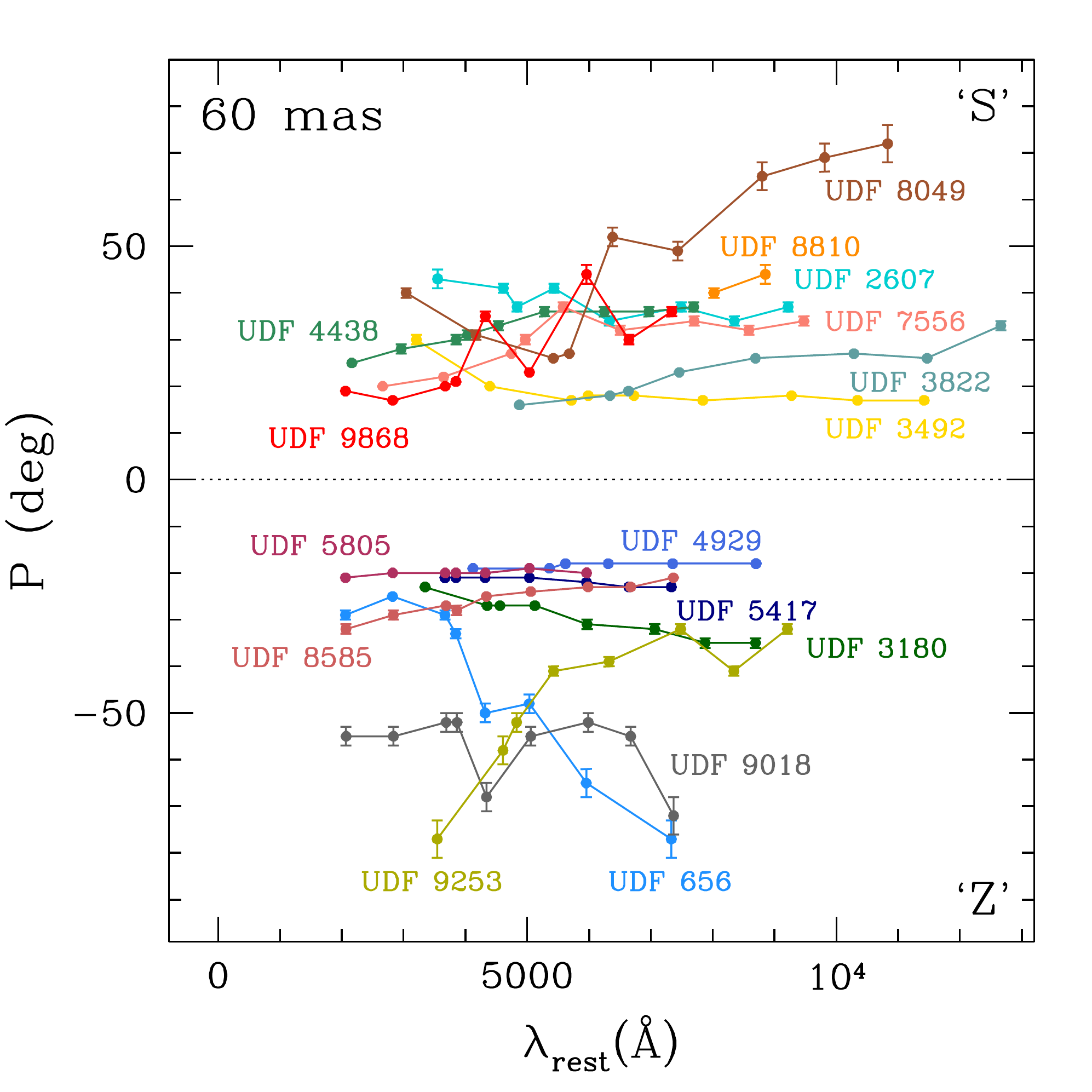}
\caption[fig12]{Same as Figure~\ref{fig11} for the 60 mas XDF images, in filters
F435W, F606W, F775W, F814W, F850LP, F105W, F125W, F140W, and F160W.
}
~\label{fig12}
\end{figure}

The results of the measured pitch angles for the 30 mas data are given in Table~\ref{tbl-4}
and plotted in Figure~\ref{fig11} as a function of rest wavelength $\lambda_{\rm rest}$(\AA).
In this work we use the following convention:
the objects with an `S' on-the-sky view have positive pitch angles ($P$),
and the objects with a `Z' on-the-sky view have negative $P$ values (see equation~\ref{eqP}).
No $P$ value is accepted when its sign does not agree with the `S' or `Z' on-the-sky view.
Similarly, the results for the 60 mas data are given in Table~\ref{tbl-5} and Figure~\ref{fig12}.

In order to quantify the correlation of $P$ with $\lambda_{\rm rest}$, we fit
a linear relation $P(\lambda_{\rm rest})= a + b\lambda_{\rm rest}$ to the values of the pitch angles
obtained from the NIR filters (see also Section~\ref{wavypatts}) for each object, individually. 
The results are given in Table~\ref{tbl-6}, where
$r_{xy}$ is the correlation coefficient~\citep{bev69},
and $\Delta\lambda_{\rm rest}$ indicates the wavelength range at rest for which the pitch angles are measured.
From these fits we can determine if $\vert P \vert$ (the absolute value
of the pitch angle) decreases for shorter wavelengths, as expected for the
colour jumps (see Section~\ref{sect_colorgrads}) of trailing spiral arms.
For most of the objects we find the expected colour jump trend, i.e., the absolute value of the pitch angle $\vert P \vert$
decreases for shorter wavelengths. However, for three objects, UDF~4929, UDF~8585, and UDF~9253,
we find the opposite, i.e., the $\vert P \vert$ value increases for shorter wavelengths.
In Table~\ref{tbl-6} we have catalogued the colour jumps for these objects as `reverse', i.e., from blue-to-red,
as opposed to `normal', i.e., from red-to-blue.
Interestingly, the three `reverse' colour jump galaxies have a `Z' on-the-sky view, but
the effect is important only for UDF~9253. In the cases of UDF 4929 and UDF 8585,
the pitch angles are almost constant with wavelength.

\subsection{Leading spiral arms?}~\label{leadingarms}

At $z\approx0$ there are practically no leading galaxies, i.e.,
with the tails of the spiral arms pointing in the direction of galactic rotation,
among the thousands known~\citep[e.g.,][]{vai08,gro08,lieb22}.
We may test whether the arms of our three `reverse' colour jump galaxies are indeed leading, using the position of the dust lanes.
For regions inside $R_{\mathrm CR}$, grand design spirals often have dust lanes along the inner edge (or the concave side) of the arms~\citep{gros99}.
Unfortunately, dust lanes are not detected in UDF~4929 and UDF~8585,
and are just barely discerned in UDF~9253, due to the limited spatial resolution.
We can also employ the method of~\citet{vai08} to determine if a galaxy is rotating clockwise
or counterclockwise. This procedure relies on the assumption that the side of a disc galaxy closer
to the observer shows more dust extinction. The sense of rotation can be inferred from
the rotation curve. To implement this method we measured the surface brightness profiles along
the minor axis of the discs in our galaxy sample. The results of this analysis are
not convincing, since we obtained nearly the same number of leading and trailing spirals.
This may indicate that the method is effective only for certain, but not all,
relative spatial distributions of dust and stars in galactic discs.
A colour asymmetry in the discs may also suggest which side is near or far from the observer.
In~\citet{mart09a}, we found that in $\sim$50\% of a sample of 31 face-on spiral galaxies of various Hubble types,
the photometric $Q$-index [$Q(rJgi) = (r-J) - \frac{E(r-J)}{E(g-i)}(g-i)$]
shows two average values, one for each half of the image. We dubbed these `Q effect' galaxies.
For this small sample it is not clear whether the disks are divided along their projected mayor axes or not.
Although the `$Q$ effect' may suggest near/far disc sides, not all the objects show it, which suggests
that the dust and stars relative distributions are not the same for all galaxies. 
The `$Q$ effect' may be a consequence of a reflection in the disc resulting in artificial asymmetries~[see the case of NGC~3521 in~\citet{zib09}].
A proposed correlation between dust attenuation and galaxy mass~\citep{lor23}
may also affect this behaviour. Having said that, any conclusion derived from colour asymmetries (even with UV filters)
would require further analysis outside the scope of this work.

\begin{figure} 
\centering
\includegraphics[width=\columnwidth]{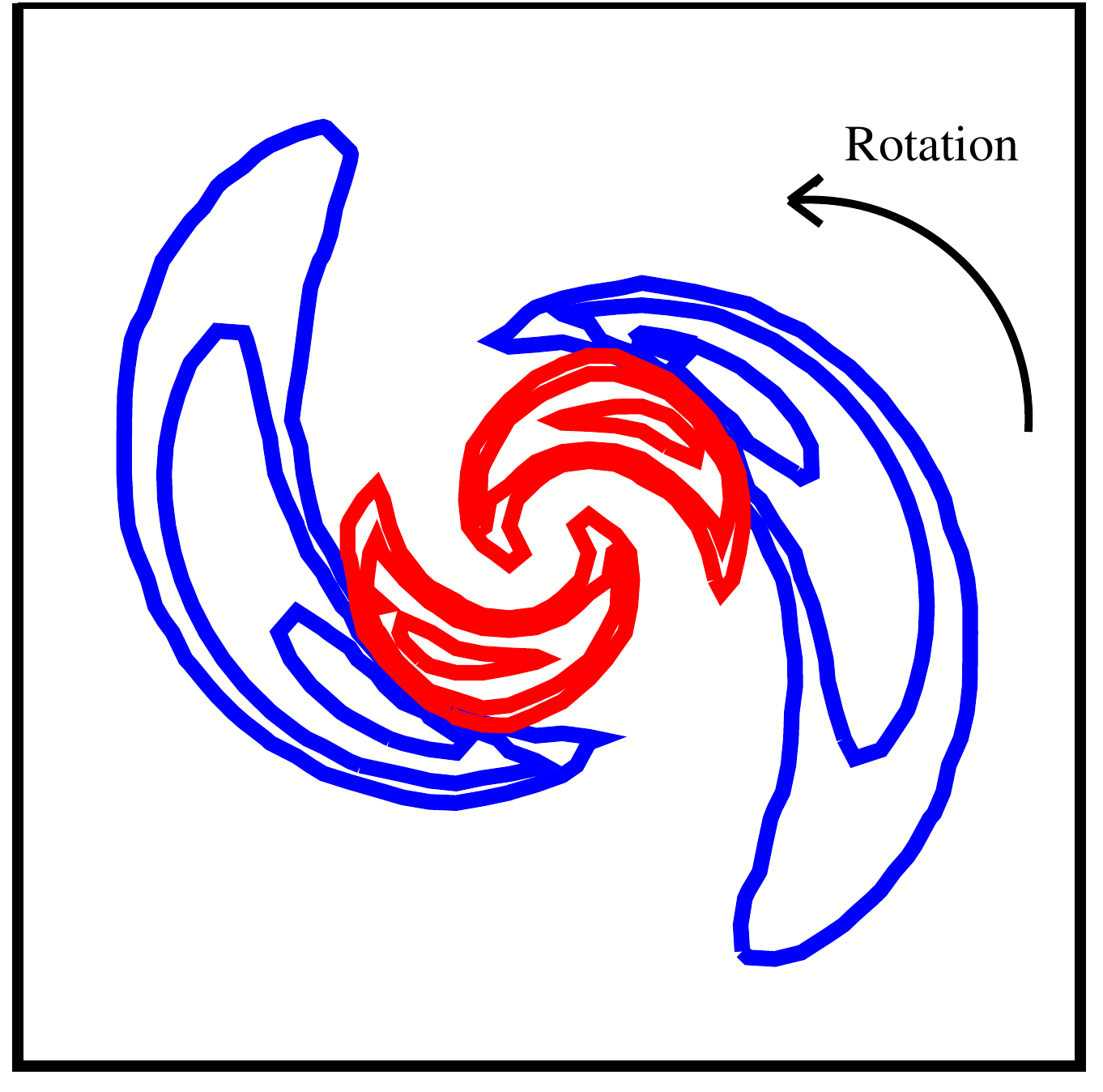}
\caption[fig13]{Contour plot of two rotating spiral patterns. The outer
pattern (blue contours) has a pitch angle of $P=24\degr$, and the inner
spiral (red contours) has $P=27\degr$.
}
~\label{fig13}
\end{figure}

\begin{figure} 
\centering
\includegraphics[width=\columnwidth]{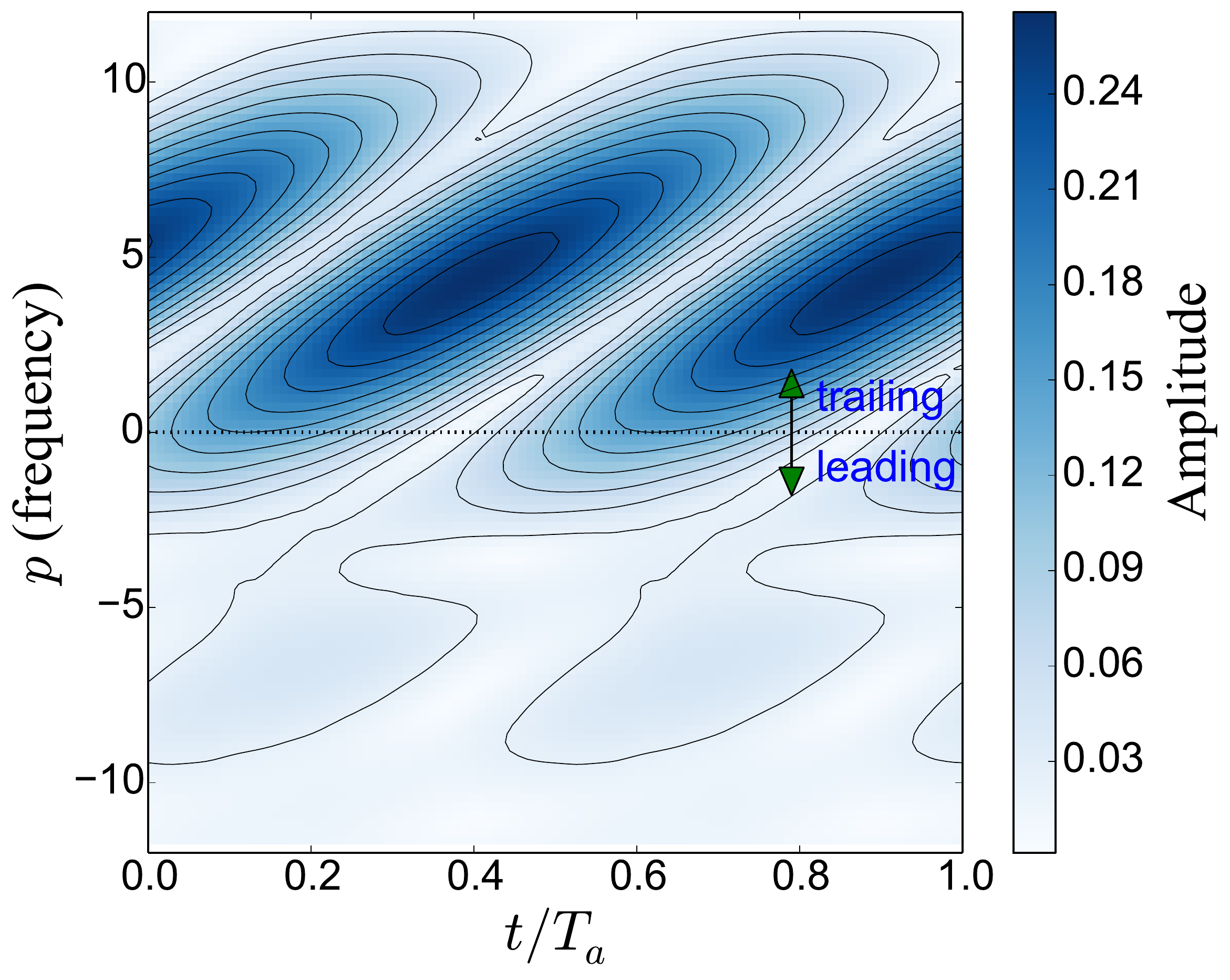}
\caption[fig14]{$A(m=2,p)$ spectra resulting from the superposition of
the two rotating spiral patterns shown in Figure~\ref{fig13}, as a function of time, $t$.
$T_{a}$ represents the time needed for the inner pattern to complete one full revolution
with respect to the outer pattern.
}
~\label{fig14}
\end{figure}

According to theoretical frameworks, the leading appearance of the spiral arms may be explained if
they are undergoing the leading phase of the swing
amplification~\citep[e.g.,][]{gol78,too81} or the WASER~\citep{mark76} mechanism.
In these frameworks, the spiral arms are expected to originate as leading waves,
evolve to amplified trailing waves, and eventually fade out.
In modern numerical simulations, `swing amplified'
spirals can be explained as the superposition of a few long-lived patterns~\citep[see, e.g.,][]{val17,sell21}.
In Figure~\ref{fig13} we show a contour plot of two uniformly-rotating spiral patterns.
If we assume that the inner spiral (red contours) rotates with a higher $\Omega_{\rm p}$,
than the outer spiral (blue contours), and use Equation~\ref{eq2DFT} to compute the $A(m=2,p)$ spectra as
a function of time, we obtain the plot in Figure~\ref{fig14}. From this graph
it is evident that the global pattern (resulting from the superposition of the two spiral patterns in Figure~\ref{fig13})
has a recurring cycle that shifts from leading ($p<0$) to trailing ($p>0$).
In order for this outcome to explain the apparently `leading'
spiral arms that we observe with the colour jumps it would be necessary for
the gas to adjust rapidly to the changes in the spiral potential~\citep[see, e.g.,][]{sell20}.
Also, the amplitude of the `leading' phase of the spiral pattern should be 
higher than the amplitude of the `trailing' phase at a given time. However,
as shown in Figure~\ref{fig14}, the amplitude of the `trailing' phase always
dominates the $A(m,p)$ spectra and, consequently,
it would be difficult to actually observe the `leading' phase.
The presence of `hidden' leading patterns can be indirectly inferred from the modulation of
the spiral arms~\citep[e.g.,][]{elm89,pue00}. These patterns will not appear directly on the images,
but they can create interference signatures on the density contours
along the arms. In our case, the apparently ‘leading’
spirals are clearly distinguishable and thus cannot be explained in the context of numerical simulations.

Finding {\it bona fide} leading spiral arms in galaxies in the distant universe may help to
better constrain some theories about the nature of spiral galaxies.
For instance, if a statistically significant number of galaxies have leading spiral arms at a certain redshift, 
there is the possibility that they are `swing amplified'~\citep{too81}.
Moreover, if leading arms are indeed due to interactions, we would expect the number of spiral galaxies with leading
arms to increase at $z=2$ and beyond, just as quasars (also attributed to interactions) have their golden age
between $z=2$ and $z=3$~\citep[e.g.,][]{shav96}. At larger redshifts, the black holes may not have had enough time to become massive.
Leading arms would only require interactions, whose number seem to diminish continuously between $z=2$ and $z=0$.
However, the question would still remain about how leading spiral arms could survive long enough to be observed as density waves.


\begin{table*}
	\centering
	\caption{Analysis parameters and measured pitch angles ($P$) for the 30 mas data.}
	\label{tbl-4}
	\begin{tabular}{ccccrrrrr} 
		\hline
		UDF-ID & I.A.~(\degr) & P.A.~(\degr) & $\Delta{R}$ & F435W~(\degr) & F606W~(\degr) & F775W~(\degr) & F814W~(\degr) & F850LP~(\degr) \\		
		\hline


656	&	52.3	&	139.3	&	(0.36-0$\farcs$90) &	-28.4	($\pm$0.9)  &	-25.4	($\pm$0.7)  &	-24.7	($\pm$0.7)  &	-24.7	($\pm$0.7)  &	-32.3	($\pm$1.1)	\\
	&		&		&	(2.98-7.46~kpc) & \\																					
2607	&	44.1	&	20.0	&	(0.78-1$\farcs$20) &	     -~-~~~~~~      &	35.3	($\pm$1.3)  &	39.6	($\pm$1.6)  &	36.6	($\pm$1.4)  &	41.2	($\pm$1.7)	\\
	&		&		&	(5.53-8.51~kpc) & \\																					
3180	&	21.3	&	159.4	&	(0.48-1$\farcs$20) &	     -~-~~~~~~      &	       -~-~~~~~~    &	-26.6	($\pm$0.8)  &	-26.6	($\pm$0.8)  &	-27.4	($\pm$0.8)	\\
	&		&		&	(3.60-8.99~kpc) & \\																					
3492	&	68.6	&	55.3	&	(1.32-2$\farcs$28) &	17.7	($\pm$0.3)  &	15.7	($\pm$0.3)  &	14.7	($\pm$0.2)  &	15.2	($\pm$0.2)  &	15.4	($\pm$0.3)	\\
	&		&		&	(6.53-11.28~kpc) & \\																					
3822	&	58.1	&	83.5	&	(0.60-1$\farcs$53) &	14.7	($\pm$0.2)  &	16.2	($\pm$0.3)  &	17.5	($\pm$0.3)  &	18.5	($\pm$0.4)  &	20.6	($\pm$0.5)	\\
	&		&		&	(2.10-5.36~kpc) & \\																					
4438	&	43.2	&	20.0	&	(0.30-1$\farcs$50) &	28.9	($\pm$0.9)  &	31.8	($\pm$1.1)  &	32.9	($\pm$1.1)  &	34.1	($\pm$1.2)  &	34.1	($\pm$1.2)	\\
	&		&		&	(2.44-12.18~kpc) & \\																					
4929	&	40.2	&	108.9	&	(0.36-1$\farcs$02) &	     -~-~~~~~~      &	-22.9	($\pm$0.6)  &	-22.9	($\pm$0.6)  &	-21.2	($\pm$0.5)  &	-20.7	($\pm$0.5)	\\
	&		&		&	(2.05-5.82~kpc) & \\																					
5417	&	33.4	&	8.7	&	(0.24-0$\farcs$78) &	     -~-~~~~~~      &	-21.1	($\pm$0.5)  &	-21.1	($\pm$0.5)  &	-21.1	($\pm$0.5)  &	-20.6	($\pm$0.5)	\\
	&		&		&	(1.99-6.46~kpc) & \\																					
5805	&	37.4	&	113.6	&	(0.18-0$\farcs$72) &	-19.3	($\pm$0.4)  &	-19.3	($\pm$0.4)  &	-19.7	($\pm$0.4)  &	-19.3	($\pm$0.4)  &	-19.3	($\pm$0.4)	\\
	&		&		&	(1.49-5.97~kpc) & \\																					
7556	&	36.9	&	175.3	&	(0.42-0$\farcs$90) &	18.9	($\pm$0.4)  &	19.7	($\pm$0.4)  &	21.2	($\pm$0.5)  &	22.2	($\pm$0.6)  &	24.7	($\pm$0.7)	\\
	&		&		&	(2.89-6.19~kpc) & \\																					
8049	&	40.7	&	8.8	&	(0.36-1$\farcs$02) &	32.8	($\pm$1.1)  &	27.9	($\pm$0.9)  &	24.8	($\pm$0.7)  &	25.5	($\pm$0.7)  &	35.2	($\pm$1.3)	\\
	&		&		&	(2.01-5.69~kpc) & \\																					
8585	&	36.0	&	24.0	&	(0.42-1$\farcs$02) &	-28.5	($\pm$0.9)  &	-26.8	($\pm$0.8)  &	-25.3	($\pm$0.7)  &	-26.1	($\pm$0.8)  &	-24.0	($\pm$0.7)	\\
	&		&		&	(3.48-8.44~kpc) & \\																					
8810	&	33.0	&	117.8	&	(0.24-0$\farcs$72) &	      -~-~~~~~~     &	       -~-~~~~~~    &	17.7	($\pm$0.4)  &	18.4	($\pm$0.4)  &	19.2	($\pm$0.4)	\\
	&		&		&	(1.77-5.31~kpc) & \\																					
9018	&	37.7	&	42.1	&	(0.24-0$\farcs$60) &	-35.0	($\pm$1.3)  &	-37.8	($\pm$1.5)  &	-37.8	($\pm$1.5)  &	-39.3	($\pm$1.6)  &	-53.2	($\pm$2.5)	\\
	&		&		&	(1.99-4.96~kpc) & \\																					
9253	&	47.8	&	55.0	&	(0.60-1$\farcs$98) &	-78.7	($\pm$3.7)  &	-65.0	($\pm$3.1)  &	-53.7	($\pm$2.5)  &	-51.3	($\pm$2.3)  &	-49.1	($\pm$2.2)	\\
	&		&		&	(4.26-14.05~kpc) & \\																					
9868	&	28.9	&	156.9	&	(1.08-1$\farcs$98) &	19.6	($\pm$0.4)  &	21.0	($\pm$0.5)  &	25.2	($\pm$0.7)  &	23.2	($\pm$0.6)  &	34.8	($\pm$1.3)	\\
	&		&		&	(8.95-16.41~kpc) & \\																					

		\hline
	\end{tabular}
\end{table*}


\begin{landscape}
\begin{table}
	\caption{Measured pitch angles ($P$) for the 60 mas data.}
	\label{tbl-5}
	\begin{tabular}{crrrrrrrrr} 
		\hline
		UDF-ID & F435W~(\degr) & F606W~(\degr) & F775W~(\degr) & F814W~(\degr) & F850LP~(\degr) & F105W~(\degr) & F125W~(\degr) & F140W~(\degr) & F160W~(\degr) \\		
		\hline


656   &       -29.2	($\pm$1.0)  &  -25.6	($\pm$0.8)  &  -29.2	($\pm$1.0)  &  -33.8	($\pm$1.3)  &  -50.6	($\pm$2.6)  &  -48.1	($\pm$2.4)  &  -65.9	($\pm$3.6)  &	       -~-~~~~~~    &  -77.4	($\pm$4.1)  \\
2607  &              -~-~~~~~~      &	43.1	($\pm$2.0)  &	41.1	($\pm$1.9)  &	37.6	($\pm$1.6)  &	41.1	($\pm$1.9)  &	34.6	($\pm$1.4)  &	37.6	($\pm$1.6)  &	34.6	($\pm$1.4)  &	37.6	($\pm$1.6)  \\
3180  &	             -~-~~~~~~      &  -23.7	($\pm$0.7)  &  -27.9	($\pm$1.0)  &  -27.9	($\pm$1.0)  &  -27.9	($\pm$1.0)  &  -31.2	($\pm$1.2)  &  -32.4	($\pm$1.3)  &  -35.2	($\pm$1.5)  &  -35.2	($\pm$1.5)  \\
3492  &	       30.1	($\pm$1.0)  &	20.9	($\pm$0.5)  &	17.1	($\pm$0.3)  &	18.2	($\pm$0.4)  &	18.6	($\pm$0.4)  &	17.9	($\pm$0.4)  &	18.2	($\pm$0.4)  &	17.1	($\pm$0.3)  &	17.5	($\pm$0.4)  \\
3822  &              -~-~~~~~~      &	16.5	($\pm$0.3)  &	18.0	($\pm$0.4)  &	19.3	($\pm$0.5)  &	23.1	($\pm$0.7)  &	26.8	($\pm$0.9)  &	27.7	($\pm$0.9)  &	26.8	($\pm$0.9)  &	33.0	($\pm$1.3)  \\
4438  &        25.9	($\pm$0.8)  &	28.6	($\pm$1.0)  &	30.8	($\pm$1.2)  &	32.0	($\pm$1.2)  &	33.2	($\pm$1.3)  &	36.1	($\pm$1.5)  &	36.1	($\pm$1.5)  &	36.1	($\pm$1.5)  &	37.6	($\pm$1.6)  \\
4929  &	             -~-~~~~~~      &  -19.8	($\pm$0.5)  &  -19.8	($\pm$0.5)  &  -18.9	($\pm$0.5)  &  -18.9	($\pm$0.5)  &  -18.9	($\pm$0.5)  &  -18.4	($\pm$0.4)  &	       -~-~~~~~~    &	        -~-~~~~~~   \\
5417  &	             -~-~~~~~~      &	     -~-~~~~~~      &  -21.3	($\pm$0.6)  &  -21.3	($\pm$0.6)  &  -21.3	($\pm$0.6)  &  -21.9	($\pm$0.6)  &  -22.5	($\pm$0.7)  &  -23.2	($\pm$0.7)  &  -23.2	($\pm$0.7)  \\
5805  &	      -21.4	($\pm$0.6)  &  -20.8	($\pm$0.6)  &  -20.8	($\pm$0.6)  &  -20.8	($\pm$0.6)  &  -20.3	($\pm$0.5)  &  -19.8	($\pm$0.5)  &  -20.8	($\pm$0.6)  &	       -~-~~~~~~    &	        -~-~~~~~~   \\
7556  &	       20.8	($\pm$0.6)  &	22.6	($\pm$0.7)  &	27.3	($\pm$0.9)  &	30.4	($\pm$1.1)  &	37.2	($\pm$1.6)  &	32.9	($\pm$1.3)  &	34.2	($\pm$1.4)  &	32.9	($\pm$1.3)  &	34.2	($\pm$1.4)  \\
8049  &	       41.0	($\pm$1.9)  &	31.8	($\pm$1.2)  &	26.6	($\pm$0.9)  &	27.5	($\pm$0.9)  &	52.5	($\pm$2.8)  &	49.8	($\pm$2.6)  &	65.3	($\pm$3.6)  &	69.0	($\pm$3.8)  &	72.9	($\pm$4.0)  \\
8585  &	      -32.8	($\pm$1.3)  &  -29.3	($\pm$1.1)  &  -27.3	($\pm$0.9)  &  -28.3	($\pm$1.0)  &  -25.5	($\pm$0.8)  &  -24.7	($\pm$0.8)  &  -23.3	($\pm$0.7)  &  -23.3	($\pm$0.7)  &  -22.0	($\pm$0.6)  \\
8810  &	             -~-~~~~~~      &         -~-~~~~~~     &         -~-~~~~~~     &          -~-~~~~~~    &          -~-~~~~~~    &         -~-~~~~~~     &            -~-~~~~~~  &	40.7	($\pm$1.9)  &	44.7	($\pm$2.2)  \\
9018  &	      -55.2	($\pm$3.0)  &  -55.2	($\pm$3.0)  &  -52.3	($\pm$2.8)  &  -52.3	($\pm$2.8)  &  -68.9	($\pm$3.9)  &  -55.2	($\pm$3.0)  &  -52.3	($\pm$2.8)  &  -55.2	($\pm$3.0)  &  -72.8	($\pm$4.0)  \\
9253  &	             -~-~~~~~~      &  -77.2	($\pm$4.1)  &  -58.8	($\pm$3.2)  &  -52.9	($\pm$2.8)  &  -41.4	($\pm$1.9)  &  -39.6	($\pm$1.8)  &  -32.2	($\pm$1.2)  &  -41.4	($\pm$1.9)  &  -32.2	($\pm$1.2)  \\
9868  &	       19.1	($\pm$0.5)  &	17.3	($\pm$0.4)  &	20.6	($\pm$0.6)  &	21.2	($\pm$0.6)  &	35.4	($\pm$1.5)  &	23.1	($\pm$0.7)  &	44.6	($\pm$2.2)  &	30.2	($\pm$1.1)  &	37.0	($\pm$1.6)  \\

		\hline
	\end{tabular}
\end{table}
\end{landscape}


\begin{table*}
	\caption{Linear fits to the pitch angles, $P(\lambda_{\rm rest})= a + b\lambda_{\rm rest}$, for filters F105W, F125W, F140W and F160W.}
	\label{tbl-6}
	\begin{tabular}{ccccccc} 
		\hline		
                UDF-ID  & View on the &      a      &  b  &       $r_{xy}$     &  $\Delta\lambda_{\rm rest}$  & Colour jump \\																																							
                     ~  &     sky     &      ~      &  ~  &       ~            &              (\AA)           & orientation \\
                \hline

656	  &    `Z'    &         11.86          &        -0.01239    	   & -0.9724  &  (5033  -  7330)  &  normal      \\
2607	  &    `S'    &         30.81	       &         0.00067           &  0.4743  &  (6330  -  9219)  &  normal	 \\
3180	  &    `Z'    &        -21.24          &  	-0.00166           & -0.9461  &  (5970	-  8695)  &  normal 	 \\
3492	  &    `S'    &         15.79          &	 0.00026           &  0.9992  &  (7841	-  9278)  &  normal	 \\
3822	  &    `S'    &         14.51          &	 0.00131           &  0.7418  &  (8694	-  12660) &  normal	 \\
4438	  &    `S'    &         32.75          &	 0.00057           &  0.7410  &  (5282	-  7693)  &  normal	 \\
4929	  &    `Z'    &        -21.33          &	 0.00033           &  1.0000  &  (7351	-  8698)  &  reverse	 \\
5417	  &    `Z'    &        -18.88          &        -0.00061	   & -0.9580  &  (5036	-  7334)  &  normal	 \\
5805	  &    `Z'    &	       -13.92          &        -0.00116           &  1.0000  &  (5036	-  5959)  &  normal	 \\
7556	  &    `S'    &         31.19          &	 0.00029           &  0.4743  &  (6506	-  9475)  &  normal	 \\
8049	  &    `S'    &         ~2.67          &	 0.00668           &  0.9569  &  (7435	-  10830) &  normal	 \\
8585	  &    `Z'    &        -30.23          &	 0.00110	   &  0.9580  &  (5055	-  7363)  &  reverse	 \\
8810	  &    `S'    &	        ~1.37 	       &	 0.00490           &  1.0000  &  (8019  -  8851)  &  normal	 \\
9018	  &    `Z'    &        -15.57          &        -0.00691	   & -0.7250  &  (5058	-  7367)  &  normal	 \\
9253	  &    `Z'    &        -47.59          &	 0.00143           &  0.3654  &  (6324	-  9211)  &  reverse	 \\
9868	  &    `S'    &         ~8.08          &	 0.00410           &  0.4387  &  (5037	-  7336)  &  normal	 \\
 
		\hline																																																		
	\end{tabular}	
\end{table*}

\subsection{Wavy patterns in the relation $\vert P \vert$ vs. $\lambda_{\rm rest}$}~\label{wavypatts}

\begin{figure*} 
\centering
\includegraphics[width=1.0\hsize]{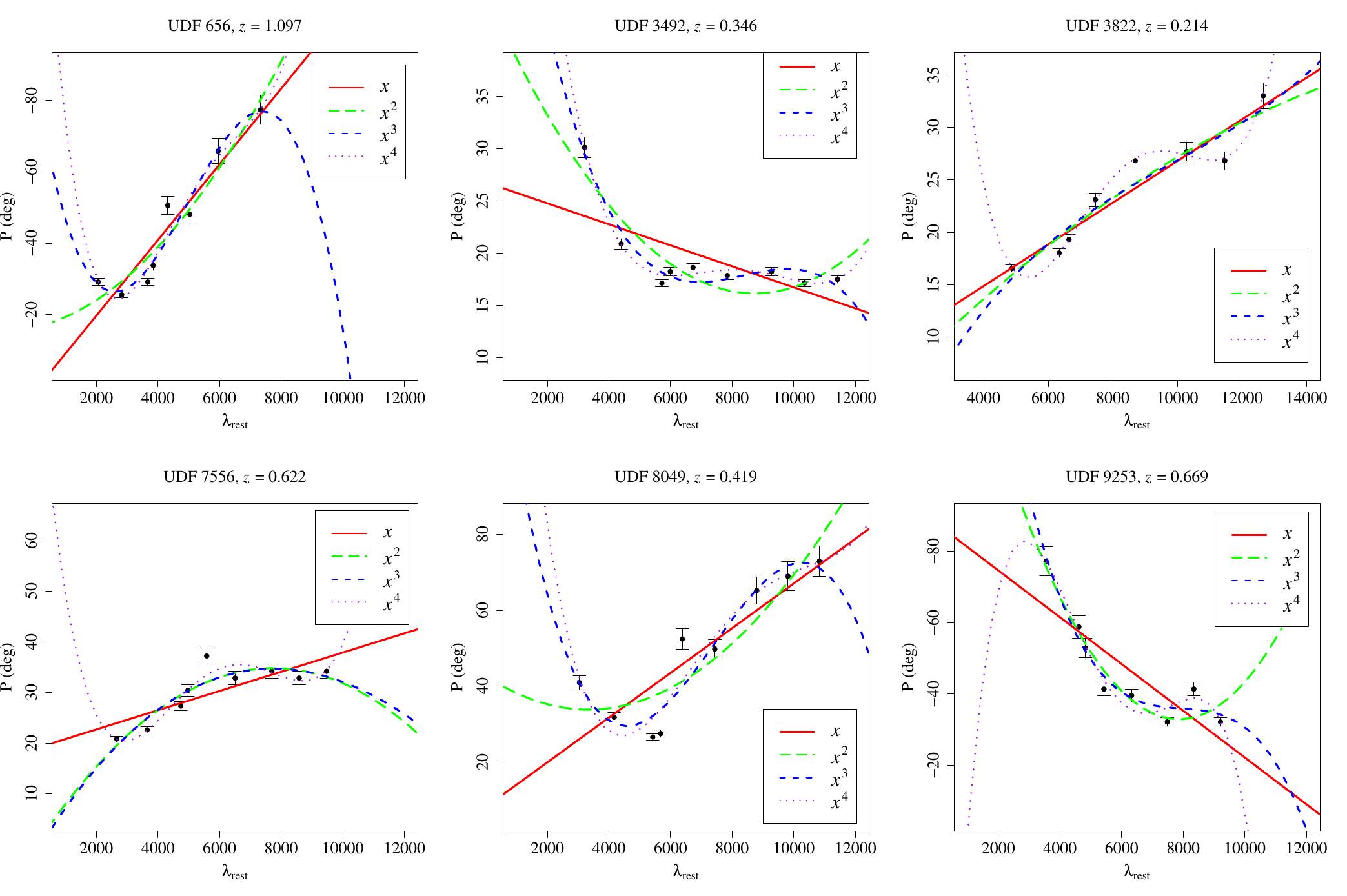}
\caption[fig15]{Polynomial fits to the pitch angle measurements, from the 60 mas data, of UDF 656, UDF 3492, UDF 3822, UDF 7556, UDF 8049, and UDF 9253.
{\it Continuous red line:} linear $x$ fit;
{\it long-dashed green line:} second order $x^2$ fit;
{\it short-dashed blue line:} third order $x^3$ fit;
{\it dotted violet line:} fourth order $x^4$ fit.
Notice the `wavy' pattern of the data.
}
~\label{fig15}
\end{figure*}

In order to quantify the relation of $\vert P \vert$ with $\lambda_{\rm rest}$ we calculated linear fits to the NIR filters only.
Although these fits represent a good approximation to the general trends,
the distributions of all data points suggest that a higher degree polynomial may be better in certain cases.
With this in mind, the relation of $\vert P \vert$ 
with $\lambda_{\rm rest}$ may be described as a `wavy' pattern for some objects.
In Figure~\ref{fig15} we show the most significant cases of `wavy' patterns
in our sample. Every fit ($x$, $x^2$, $x^3$ and $x^4$) gives the same weight to all data points.

\begin{figure*} 
\centering
\includegraphics[width=1.0\hsize]{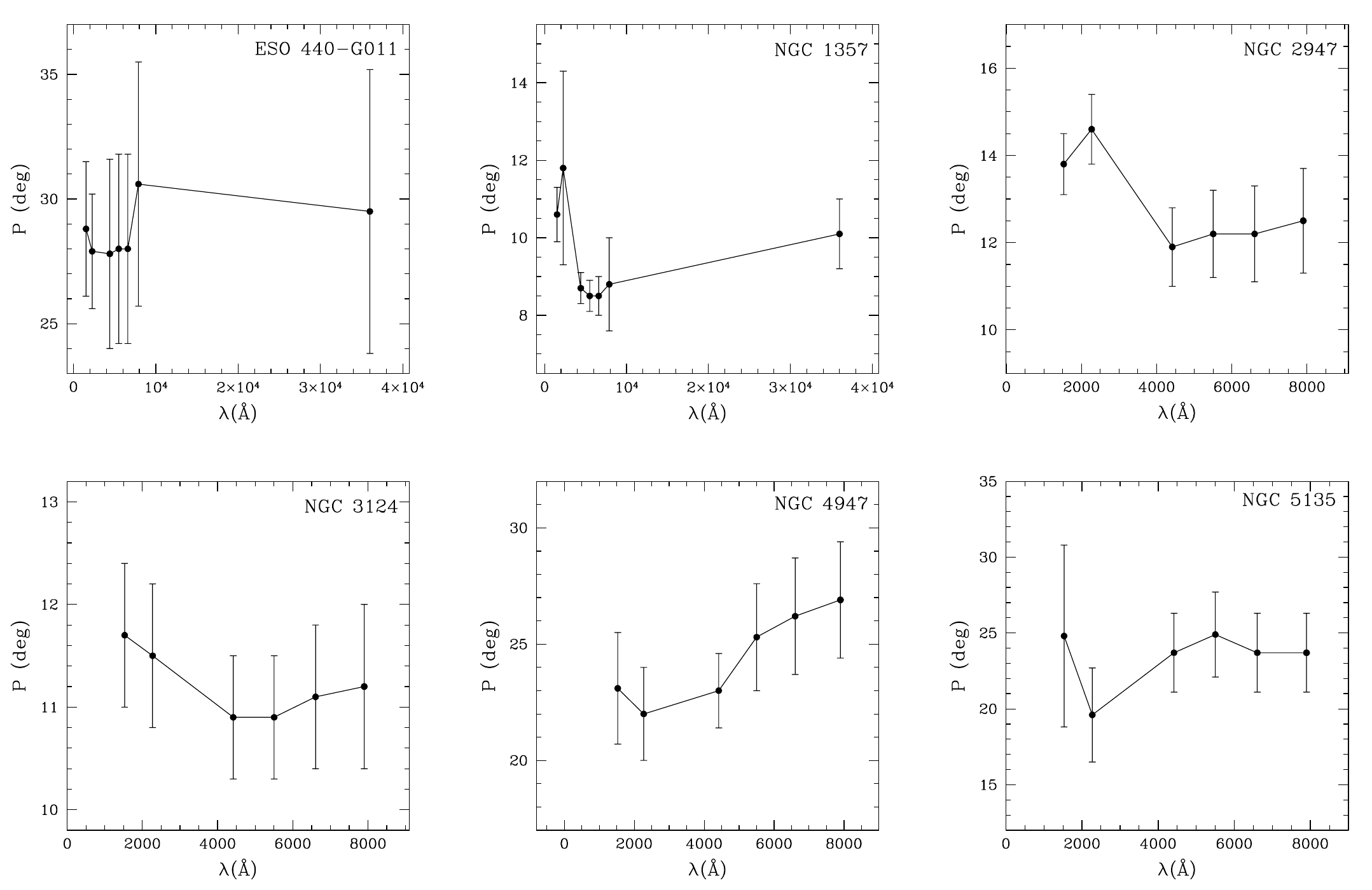}
\caption[fig16]{Examples of `wavy' patterns detected in the $P$ measurements of ESO 440-GO11, NGC 1357, 2947, 3124, 4947,
and NGC 5135, performed by~\citet{yu18}.
}
~\label{fig16}
\end{figure*}

In the case of spirals with `normal' colour jumps (see Table~\ref{tbl-6}), the first aspect to notice
is that the $\vert P \vert$ values at shorter wavelengths tend to
switch, i.e., the $\vert P \vert$ value increases for shorter wavelengths.
This may be explained if the classic age/colour gradients (see Figure~\ref{fig1}, left panel)
begins to dominate at places where the optical depth $\tau<1$, although still close to the dust lane and to SF-arm.
In other cases, $P$ switches again at longer wavelengths.
The pitch angle of the dust lanes, $P_{\rm shock}$,
can be smaller or equal to the pitch angle of the density wave traced by old stars, $P_{\rm pot}$~\citep{gitt04}.
Thus, the wavy pattern may be explained as a combination of various factors, including
dust lanes and colour jumps at longer $\lambda_{\rm rest}$, as well as
classic colour gradients at shorter wavelengths. However, the colour jump
dominates most of the $P$ measurements. This scenario may also explain the `reverse' colour jumps discussed earlier.
The `wavy' pattern was also detected in the pitch angle measurements of spiral arms in galaxies at $z=0$ by~\citet{yu18};
40\% of their objects (see Figure~\ref{fig16}) with optical, $NUV$ and $FUV$ data present this phenomenon.
This conclusion may also explain the findings of other authors for nearby galaxies~\citep{pou16,mill19,abd22},
whose $P$ measurements of the colour jump show smaller angles in the red than in the blue, as expected for the classic gradients.
Unfortunately, the aforementioned works do not provide values for $\Delta{R}$, and hence it is difficult
to reproduce their results, and to compare them with those of other authors.

\section{Conclusions}~\label{conclu}

By measuring the pitch angles of distant galaxies in the Hubble XDF~\citep{illi13},
we have found evidence of colour jumps (red-to-blue) across the spiral arms
of disc galaxies at $z>0$. Similarly to their local
counterparts, most of these galaxies exhibit trailing spiral arms.
However, for three out of sixteen objects the colour jump measurements
are found to be reverse, i.e., blue-to-red, and may indicate leading spiral arms.
Also, the behaviour of the pitch angle as a function of rest wavelength is not linear, but resembles a wavy pattern.
We presume that the reverse colour jump and the wavy pattern may be
due to a combination of classic age/colour gradients
and colour jumps across the spiral arms, together with extinction due to the dust lanes.

These results indicate that colour jumps can be observed across the spiral arms of galaxies
between redshifts of $z=0-1$, i.e., for the last $\sim8$ Gyr.
The onset of spiral structure may occur within the redshift
range of $1.5<z<3$~\citep{elm14,hod19,marg22}.
Spiral galaxies at redshifts $z>2$ may be common;
for instance, we have the spiral galaxies HDFX~28 at $z=2.011$~\citep{daw03},
Q2343-BX442 at $z=2.18$~\citep{law12},
A1689B11 at $z=2.54$~\citep{yua17},
A2744-DSG-$z$3 at $z=3.059$~\citep{wu22},
and even BRI~1335-0417 at redshift $4.41$~\citep{tsu21}.
Besides, there is a predominance of disc galaxies for $z>1.5$~\citep{ferr22,robe22},
and a growing evidence of cold rotating discs at $z>4$~\citep[e.g.,][]{deb14,riz20,kre22,tok22}.
The results of this paper stress the importance of future studies
with larger samples of distant disc galaxies in order to unveil the origin of spiral
structure in the universe.

\section*{Acknowledgements}
We acknowledge the reviewer for important comments and suggestions.
EMG acknowledges support through the `Investigadores por M\'exico' (formerly C\'atedras CONACYT) program.
RAGL acknowledges the financial support of CONACyT, Mexico, through project A1-S-8263.

\section*{Data availability}

Based on observations made with the NASA/ESA Hubble Space Telescope,~\url{https://archive.stsci.edu/prepds/xdf/}, 
and obtained from the Hubble Legacy Archive,~\url{https://hla.stsci.edu/hlaview.html};
which is a collaboration between the Space Telescope Science Institute (STScI/NASA),
the Space Telescope European Coordinating Facility (ST-ECF/ESA) and the Canadian Astronomy Data Centre (CADC/NRC/CSA).

\noindent Based on data obtained from the ESO Science Archive Facility,
\url{http://archive.eso.org/wdb/wdb/adp/phase3_spectral/form?collection_name=MUSE};
under request numbers 647703, 647706, 647710, 647712, 647716, 647719, and 647724 `PHASE3'.








\appendix

\section{Configuration used in {\tt{CIGALE}}}~\label{appA}

In this section we list the input parameters configuration used in {\tt{CIGALE}}~\citep[cf.][]{boq19} to fit the HUDF galaxies
with~\citet{rafel15} photometry.

\begin{itemize}

\item  {\tt{sfh2exp}} module

    \begin{itemize}
    \item 
    tau\textunderscore main = 1000, 2000, 3000, 5000, 7000 (Myr)
    \item 
    tau\textunderscore burst = 10, 20, 30, 50, 70 (Myr)
    \item 
    f\textunderscore burst = 0.01, 0.1, 0.2, 0.4, 0.6
    \item 
    age = 3000, 5000, 7000, 9000, 11000, 13000 (Myr)
    \item 
    burst\textunderscore age = 200, 600, 800, 1000 (Myr)
    \item 
    sfr\textunderscore 0 = 1.0 (M$_{\sun}$ yr$^{-1}$)
    \item 
    normalise = True
    \end{itemize}

\item  {\tt{bc03}} module

    \begin{itemize}
    \item 
    imf = 1 (Chabrier)
    \item 
    metallicity = 0.0001, 0.0004, 0.004, 0.008, 0.02, 0.05
    \item 
    separation\textunderscore age = 10 (Myr)
    \end{itemize}

\item  {\tt{dustatt\textunderscore modified\textunderscore CF00}} module

    \begin{itemize}
    \item 
    Av\textunderscore ISM = 1, 2, 3 ,4
    \item 
    mu = 0.1, 0.3, 0.6, 0.9
    \item 
    slope\textunderscore ISM = -0.7
    \item 
    slope\textunderscore BC = -1.3
    \end{itemize}

\end{itemize}

\noindent With this configuration {\tt{CIGALE}} computes $4.32\times10^6$ models,
i.e., every combination of the input parameters.

\bsp	
\label{lastpage}
\end{document}